\documentclass[iop]{emulateapj}
\usepackage{natbib,graphicx,epstopdf,amsmath}

\newcommand{\logr}{log(\ensuremath{R'_{\mbox{\scriptsize HK}}})}

\shorttitle{3.6 and 4.5~\micron~Phase Curves for HD~189733b}
\shortauthors{Knutson et al.}\def\simgr{\,\hbox{\hbox{$ > $}\kern -0.8em \lower 1.0ex\hbox{$\sim$}}\,}
\def\simle{\,\hbox{\hbox{$ < $}\kern -0.8em \lower 1.0ex\hbox{$\sim$}}\,}

\begin{document}

\title{3.6 and 4.5~\micron~Phase Curves and Evidence for Non-Equilibrium Chemistry in the Atmosphere of Extrasolar Planet HD~189733b} 

\author{
Heather A. Knutson\altaffilmark{1,2}, Nikole Lewis\altaffilmark{3}, Jonathan J. Fortney\altaffilmark{4}, Adam Burrows\altaffilmark{5}, Adam P. Showman\altaffilmark{3}, Nicolas B. Cowan\altaffilmark{6}, Eric Agol\altaffilmark{7}, Suzanne Aigrain\altaffilmark{8}, David Charbonneau\altaffilmark{9}, Drake Deming\altaffilmark{10}, Jean-Michel D\' esert\altaffilmark{9}, Gregory W. Henry\altaffilmark{11}, Jonathan Langton\altaffilmark{12}, Gregory Laughlin\altaffilmark{4}
}

\altaffiltext{1}{Division of Geological and Planetary Sciences, California Institute of Technology, Pasadena, CA 91125, USA} 
\altaffiltext{2}{hknutson@caltech.edu}
\altaffiltext{3}{Dept. of Planetary Sciences and Lunar and Planetary Laboratory, University of Arizona, Tucson, AZ 85721, USA}
\altaffiltext{4}{Dept. of Astronomy and Astrophysics, University of California, Santa Cruz, CA 95064, USA}
\altaffiltext{5}{Dept. of Astrophysical Sciences, Princeton University, Princeton, NJ 08544, USA}
\altaffiltext{6}{CIERA, Northwestern University, Evanston, IL 60208}
\altaffiltext{7}{Dept. of Astronomy, University of Washington, Seattle, WA 98195, USA}
\altaffiltext{8}{Sub-dept. of Astrophysics, Dept. of Physics, University of Oxford, Oxford OX1 3RH, UK}
\altaffiltext{9}{Harvard-Smithsonian Center for Astrophysics, 60 Garden St., Cambridge, MA 02138, USA}
\altaffiltext{10}{Dept. of Astronomy, University of Maryland, College Park, MD, 20742, USA}
\altaffiltext{11}{Center of Excellence in Information Systems, Tennessee State University, 3500 John A. Merritt Blvd., Box 9501, Nashville, TN 37209}
\altaffiltext{12}{Department of Physics, Principia College, 1 Maybeck Place, Elsah, IL 62028, USA}

\begin{abstract}

We present new, full-orbit observations of the infrared phase variations of the canonical hot Jupiter HD 189733b obtained in the 3.6 and 4.5~\micron~bands using the \emph{Spitzer Space Telescope}.  When combined with previous phase curve observations at 8.0 and 24~\micron, these data allow us to characterize the exoplanet's emission spectrum as a function of planetary longitude and to search for local variations in its vertical thermal profile and atmospheric composition.  We utilize an improved method for removing the effects of intrapixel sensitivity variations and robustly extracting phase curve signals from these data, and we calculate our best-fit parameters and uncertainties using a wavelet-based Markov Chain Monte Carlo analysis that accounts for the presence of time-correlated noise in our data.  We measure a phase curve amplitude of $0.1242\%\pm0.0061\%$ in the 3.6~\micron~band and $0.0982\%\pm0.0089\%$ in the 4.5~\micron~band, corresponding to brightness temperature contrasts of $503\pm21$~K and $264\pm24$~K, respectively.  We find that the times of minimum and maximum flux occur several hours earlier than predicted for an atmosphere in radiative equilibrium, consistent with the eastward advection of gas by an equatorial super-rotating jet.  The locations of the flux minima in our new data differ from our previous observations at 8~\micron, and we present new evidence indicating that the flux minimum observed in the 8~\micron~is likely caused by an over-shooting effect in the 8~\micron~array.  We obtain improved estimates for HD~189733b's dayside planet-star flux ratio of $0.1466\%\pm0.0040\%$ in the 3.6~\micron~band and $0.1787\%\pm0.0038\%$ in the 4.5~\micron~band, corresponding to brightness temperatures of $1328\pm11$~K and $1192\pm9$~K, respectively; these are the most accurate secondary eclipse depths obtained to date for an extrasolar planet.  We compare our new dayside and nightside spectra for HD~189733b to the predictions of 1D radiative transfer models from \citet{burrows08}, and conclude that fits to this planet's dayside spectrum provide a reasonably accurate estimate of the amount of energy transported to the night side.  Our 3.6 and 4.5~\micron~phase curves are generally in good agreement with the predictions of general circulation models for this planet from \citet{showman09}, although we require either excess drag or slower rotation rates in order to match the locations of the measured maxima and minima in the 4.5, 8.0, and 24~\micron~bands.  We find that HD~189733b's 4.5~\micron~nightside flux is $3.3\sigma$ smaller than predicted by these models, which assume that the chemistry is in local thermal equilibrium.  We conclude that this discrepancy is best-explained by vertical mixing, which should lead to an excess of CO and correspondingly enhanced 4.5~\micron~absorption in this region.  This result is consistent with our constraints on the planet's transmission spectrum, which also suggest excess absorption in the 4.5~\micron~band at the day-night terminator.

\end{abstract}


\section{Introduction}\label{intro}

Observations of eclipsing extrasolar planetary systems, in which the planet periodically passes in front of and then behind its host star, have proven to be a powerful diagnostic tool for studies of exoplanetary atmospheres.  Because the probability of a transiting geometry scales as $R_{\star}/a$ where $R_{\star}$ is the radius of the host star and $a$ is the planet's semi-major axis, the majority of currently known transiting planet systems have orbital periods of just a few days.  At these distances, the time scale for the planet to achieve synchronous rotation is short compared to the typical ages of the systems, leading to the prediction that a majority of these transiting planets should be tidally locked \citep{bodenheimer01}.  In this paper we focus on the class of gas giant planets known as ``hot Jupiters", which typically have orbital periods on the order of $1-3$ days and atmospheric temperatures ranging between $1000-3000$~K.  

One fundamental question for these planets is what fraction of the incident flux absorbed on the planet's day side is subsequently transported around to the night side.  Atmospheric circulation models predict that these planets should develop a broad superrotating (eastward) equatorial jet that circulates energy between the day and night sides \citep[e.g.,][]{showman02,showman09,showman11,langton08,dobbs10,heng11,rauscher11}.  Depending on the relative strengths of these winds, these planets could exhibit large gradients in both temperature and composition between the two hemispheres.  We can constrain the efficiency of the day-night circulation by measuring changes in the infrared brightness of the planet as a function of orbital phase; the day-night brightness contrast can then be translated into a day-night temperature contrast.  There are currently well-characterized phase curves published for seven planets, including $\upsilon$~And b \citep{harrington06,crossfield10}, HD~189733b \citep{knutson07,knutson09a}, HD 149026b \citep{knutson09c}, HD 80606b \citet{laughlin09}, HAT-P-7b \citep{borucki09,welsh10}, CoRoT-1b \citep{snellen09}, and WASP-12b \citep{cowan12}, with more sparsely sampled phase curves for three additional planets (51 Peg b, HD 209458b, and HD 179949b) from \citet{cowan07}.  These data indicate that hot Jupiters display a diversity of circulation patterns, ranging from relatively small day-night temperature gradients (e.g., HD 189733b) to large temperature gradients (e.g., WASP-12b).

Of these four systems, HD~189733b stands out both as having the best-characterized phase variation, and also as the only system with phase curve observations at more than one wavelength.  We know more about this planet's atmosphere than that of any other extrasolar planet; key results include the detection of a high-altitude haze  \citep{pont08,sing11} and sodium absorption \citep{redfield08,huitson12} in its visible-light transmission spectrum, as well as more controversial detections of methane, carbon monoxide, and water absorption in its infrared transmission spectrum \citep{swain08,sing09,desert09,gibson11a,gibson11} and carbon dioxide absorption in its dayside emission spectrum \citep{swain09}.  Several recent ground-based studies \citep{swain10,waldmann12} have also reported detections of methane emission from the planet's day side, although these results have been the subject of some debate \citep{mandell11}.  HD 189733b's dayside emission spectrum has also been characterized in the near- and mid-infrared using both \emph{Spitzer} IRAC photometry \citep{charbonneau08} and IRS spectroscopy  \citep{grillmair08}, with additional constraints on its variability in the 8~\micron~IRAC band from \citet{agol10}.  

Despite the extent of the data available for this planet, there are still a number of open questions regarding the properties of its atmosphere.  The single largest outstanding question centers on the issue of whether or not the chemistry is in equilibrium \citep[e.g.,][]{moses11,visscher11}; it is likely that the atmospheric circulation plays an important role in shaping this chemistry \citep[e.g.,][]{cooper06}.  Although we have observational constraints on relative abundances for the planet's day side and the day-night terminator, we know very little about the properties of HD~189733b's night side.  In this paper we present new full-orbit phase curve observations for HD~189733b in the 3.6 and 4.5~\micron~bands obtained with the \emph{Spitzer Space Telescope} during its extended warm mission.  We combine these data with previous observations at 8.0 and 24~\micron~to provide the first detailed characterization of its emission spectrum as a function of orbital phase.  Our data include two secondary eclipses and one transit in each band, which we use to derive improved estimates for the planet's orbital ephemeris and dayside emission spectrum.  

\section{Observations}\label{obs}

\begin{figure}
\epsscale{1.2}
\plotone{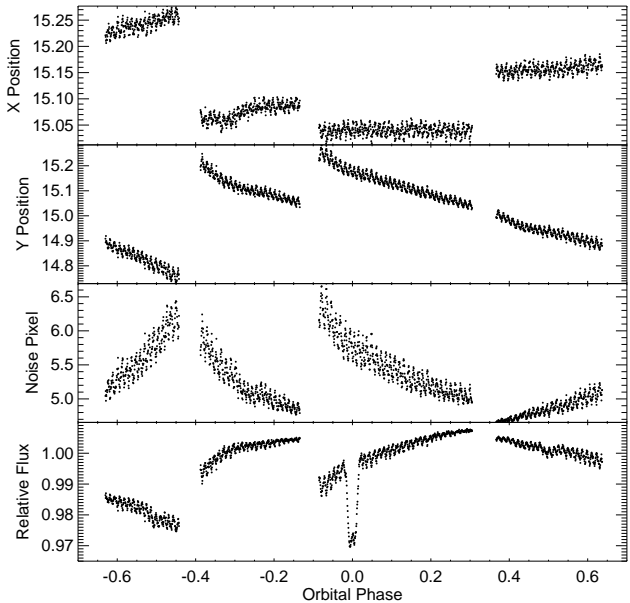}
\caption{Measured $x$ positions (top panel), $y$ positions (upper middle panel), and noise pixel values (lower middle panel) as a function of orbital phase for the 3.6~\micron~phase curve observations.  Data has been binned into two minute intervals. The raw photometry is shown in the bottom panel.  Gaps in data are due to spacecraft downlinks, and the offset in positions between downlinks is a result of the need to re-acquire the star after each downlink.}  
\label{ch1_raw}
\end{figure}

\begin{figure}
\epsscale{1.2}
\plotone{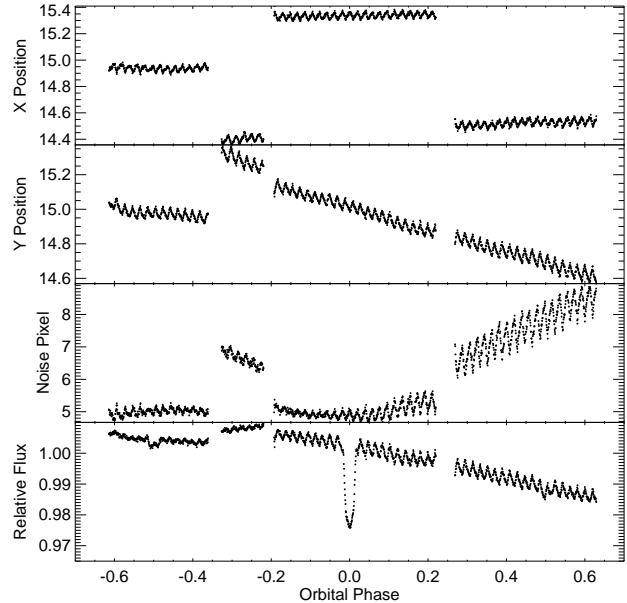}
\caption{Measured  $x$ positions (top panel), $y$ positions (upper middle panel), and noise pixel values (lower middle panel) as a function of orbital phase for the 4.5~\micron~phase curve observations.  The raw photometry is shown in the bottom panel; see Fig. \ref{ch1_raw} for a complete description.}
\label{ch2_raw}
\end{figure}

We analyze two full-orbit phase curves for HD 189733b, obtained in the 3.6 and 4.5~micron bands using the IRAC instrument \citep{faz04} on the \emph{Spitzer Space Telescope} \citep{wern04}.   The 3.6~\micron~observation was obtained on UT 2010 Dec $27-29$ and has a total duration of 68.5 hours, corresponding to 1,722,624 images.  The 4.5~\micron~observation was obtained on UT 2009 Dec $22-24$ and has a total duration of 66.7 hours, corresponding to 1,711,872 images.  Data in both bandpasses were obtained in subarray mode with 0.1~s exposures in order to avoid saturation and to minimize the total data volume.  As a result of the limited on-board memory, both observations required three breaks for downlinks with a duration of $1-2$~hours per downlink.  In order to downlink data the spacecraft must point towards the earth and then reacquire the target, which produced offsets of up to half a pixel in the star's position after each downlink (Fig. \ref{ch1_raw} and \ref{ch2_raw}).

We calculate the BJD\_UTC values at mid-exposure for each image using the MBJD\_OBS keyword in the image headers.  Each set of 64 images obtained in subarray mode comes as a single FITS file with a time stamp corresponding to the start of the first image; we calculate the time stamps for individual images assuming uniform spacing and using the difference between the AINTBEG and ATIMEEND headers, which record the start and end of the 64-image series.  \citet{eastman10} further advocate a conversion from UTC to TT timing standards, which includes a more consistent treatment of leap seconds.  We note that for the dates spanned by these observations the conversion from BJD\_UTC to BJD\_TT simply requires the addition of 66.184~s, and we proceed using BJD\_UTC in order to ensure consistency with other studies of this planet.

\subsection{Photometry}\label{phot}

Subarray images have dimensions of $32\times32$ pixels, making it challenging to estimate the sky background independent of contamination from the wings of the star's point spread function.  We choose to exclude pixels within a radius of 12 pixels of the star's position, as well as the 13th-16th rows and the 14th-15th columns, where the stellar point spread function extends close to the edge of the array.  We also exclude the top (32nd) row of pixels, which have values that are consistently lower than those for the rest of the array.  Lastly, we mask out a $4\times3$ pixel box centered on the position of HD 189733b's M star companion \citep{bakos06a}.  We then iteratively trim 3$\sigma$~outliers from the remaining subset of pixels, create a histogram of the remaining values, and fit a Gaussian to this histogram to determine the sky background for each image.  We find that the background contributes $0.2$\% of the total flux at 3.6~\micron~and $0.1$\% of the total flux at 4.5~\micron.

We determine the position of the star on the array in each image using flux-weighted centroiding  \citep[e.g.,][]{knutson08,charbonneau08}.  We also tried fits using Gaussian position estimates \citep[e.g.,][]{stevenson10,agol10}, but found that this method results in an inferior correction for the intrapixel sensitivity variations in our data.   Before estimating the position of the star on the array we first correct for transient hot pixels in each set of 64 images by replacing outliers more than $3\sigma$ away from the median flux at a given pixel position with this median value.  We then subtract the best-fit background flux from each image and iteratively calculate the flux-weighted centroid for a circular region with a radius of either 4.0 (3.6~\micron) or 3.5 (4.5~\micron) pixels centered on the estimated position of the star.  Increasing or decreasing the size of this region does not significantly alter the time series but does result in a slightly higher scatter in the normalized light curve.  We also estimate the width of the stellar point spread function in each image by calculating a quantity known as the noise pixel parameter \citep{mighell05,lewis12}, which is defined in Section 2.2.2 (``IRAC Image Quality") of the IRAC instrument handbook as:

\begin{equation}\label{eq1}
\tilde{\beta} =  \frac{\left(\sum_i I_i\right)^2}{\sum_i I_i^2}
\end{equation}
where $I_i$ is the measured intensity in the $i^{th}$ pixel.  This noise pixel parameter is equal to one over the sharpness parameter $S_1$, first introduced by \citet{muller74}, and proportional to the FWHM of the stellar point spread function squared \citep{mighell05}.  We find that we obtain the best results in both bands when we calculate $\tilde{\beta}$ using a circular aperture with a radius of 4.0 pixels.

We calculate the flux in each image using aperture photometry with either a fixed radius ranging between 2.0-5.0 pixels in 0.1 pixel steps or a time-varying radius equal to the square root of the noise pixel parameter $\tilde{\beta}$ with either a constant scaling factor ranging between $0.8-1.2$ or a constant offset between $-0.4$ and $+0.4$ pixels.  We optimize our choice of aperture individually for each of the four data segments defined by the downlink breaks, as the star falls on a different region of the pixel in each section.  We find that for the 3.6~\micron~array we obtain the best results for apertures with radii equal to $\sqrt(\tilde{\beta}$)+$[0.1,0.2,0.2,0.2]$ pixels respectively in our four segments; this corresponds to a range of $2.3-2.6$ pixels with a median value of 2.5 pixels.  In the 4.5~\micron~band we prefer a fixed aperture with a radius of 2.3 pixels in all segments.  We obtain consistent results for the best-fit phase curve and eclipse parameters in both bands over a range of either fixed (4.5~\micron) or time-varying (3.6~\micron) aperture sizes, albeit with a larger standard deviation and correspondingly higher uncertainties for the less optimal apertures.  We remove outliers from our final light curves using a moving median filter in flux with a width of 50 points, where we discard points that lie more than $5\sigma$ away from the median flux values.  This corresponds to 0.008\% of our data at 3.6~\micron~and 0.025\% at 4.5~micron.

In addition to the intrapixel sensitivity variations described in \S\ref{intrapixel}, our data also exhibit a short duration, ramp-like behavior similar to that observed in the IRAC 8.0~\micron~array \citep[e.g.][]{knutson07,knutson09c,agol10} at the start of each phase curve observation and again after each downlink break.  We find that the amplitude of this ramp decreases in each successive segment, with the largest ramp at the start of the first segment.  This effect may be due to charge-trapping in the array, as has been suggested for longer wavelengths \citep[e.g.,][]{knutson07}, or it may be related to a settling of the telescope at a new pointing.  This effect has an asymptotic shape and generally converges to a constant value on time scales of an hour or less; we elect to trim the first 60 minutes of data at the start of the 3.6~\micron~phase curve observation and the first 30 minutes of data at the start of the 4.5~\micron~phase curve observation, repeating this trim after each downlink break.  As before, we select our trim intervals in order to minimize the standard deviation of the residuals from the best-fit solution.  Such trimming is standard practice for \emph{Spitzer} secondary eclipse observations in these bands, which typically have a duration of less than eight hours \citep[e.g.][]{knutson09b,todorov10,fressin10,odonovan10,deming11}.  We also tried including exponential functions at the start of each data segment in our fits to the 3.6~\micron~phase curve, which displayed a stronger ramp than our 4.5~\micron~data, but we found that these functions did not significantly improve the quality of the fits, nor did they change the best-fit phase curve amplitude and the times of minimum and maximum flux.  

\subsection{Correction for Intrapixel Sensitivity Variations}\label{intrapixel}

Fluxes measured at these two wavelengths show a strong correlation with the changing position of the star on the array.  This effect is due to a well-documented intra-pixel sensitivity variation \citep[e.g.,][]{reach05,charbonneau05,charbonneau08,morales06,knutson08}, in which the sensitivity of an individual pixel varies between the center and the edge.  We find that quadratic functions of $x$ and $y$ position, which are commonly used to correct shorter ($<10$~hour) transit and secondary eclipse observations, are inadequate to describe the pixel response over the range of positions spanned by our phase curve observations.  We instead correct for these sensitivity variations by approximating the star as a point source on the array and using the measured fluxes as a function of $x$ and $y$ position to create a map of the pixel response.  Our approach is based on the method described in \citet{ballard10}, but includes a number of changes that improve the performance and reduce computational overheads for our long light curves (75 hours, vs. approximately 20 hours in Ballard et al.).  We calculate the effective pixel sensitivity at a given position as follows:

\begin{align} \label{eq2}
F_{meas,j} & = & F_{0,j} \sum_{i=0}^{n} e^{-\left(x_i-x_j\right)^2/2 \sigma_{x,j}^2} \nonumber \\
& & \times e^{-\left(y_i-y_j\right)^2/2 \sigma_{y,j}^2} \times e^{-\left(\sqrt{\tilde{\beta}_i}-\sqrt{\tilde{\beta}_j}\right)^2/2 \sigma_{\sqrt{\tilde{\beta},j}}^2}
\end{align}
where $F_{meas,j}$ is the measured flux in the $j$th image, $F_{0,j}$ is the intrinsic flux, $x_j$, $y_j$, and $\tilde{\beta}_j$ are the measured $x$ position, $y$ position, and noise pixel values, and $\sigma_{x,j}$, $\sigma_{y,j}$, and $\sqrt{\sigma_{\tilde{\beta},j}}$ are the the standard deviations of the $x$, $y$, and $\sqrt{\tilde{\beta}}$ vectors over the full range in $i$.  In this case the standard deviation does not represent the uncertainty on the measured positions but instead reflects the relative range in position spanned by the points in our vector.  By using this parameter to determine our smoothing width, we effectively implement an adaptive smoothing scheme with a smaller spatial resolution in regions with dense sampling and a larger resolution in regions where the pixel is more sparsely sampled. The $i$ index sums over the nearest 50 neighbors with distance $d_{i,j}$ defined as follows:

\begin{equation}\label{eq3}
d_{i,j} = \left(x_i-x_j\right)^2+  \left(\frac{y_i-y_j}{b}\right)^2+ \left(\sqrt{\tilde{\beta}_i}-\sqrt{\tilde{\beta}_j}\right)^2
\end{equation}
where $b$ is an empirically determined scaling constant that allows more weight to be placed on either the $x$ or $y$ separation.  Both $b$ and the total number of points $n$ used in Eq. \ref{eq2} are fixed at the values that produce the lowest standard deviation in the residuals from the best-fit solution.  In this case we obtain the best results when $b=0.8$ in the 3.6~\micron~band, while we prefer $b=1.0$ in the 4.5~\micron~band.  We find that increasing the number of nearest neighbors used in the weighted average from 50 to 400 reduces the relative scatter in the final unbinned residuals by less than 1\% and increases the computation time by a factor of six, making our Markov chain fits computationally intractable.  

We calculate the intrapixel sensitivity correction using the residuals after the best fit functions for the phase curve (\S\ref{phase_curve_funct}), transit and secondary eclipses (\S\ref{transit_funct}), and stellar activity (\S\ref{star_spots}) have been removed in order to ensure that we do not inadvertently remove part of these astrophysical signals.  We re-calculate this correction at each step in our fits, allowing the pixel map to trade off against changes in our nominal astrophysical functions.  We find that this iterative approach results in a unique best-fit solution despite the large amplitudes of the intrapixel sensitivity variations, as improved estimates for the astrophysical parameters result in a more accurate pixel map, which in turn results in a lower scatter in the residuals after both the astrophysical signals and the intrapixel sensitivity variations have been removed.  It is worth noting, however, that the success of this approach depends on our ability to generate accurate models for the astrophysical signals in our data.  

\subsection{Transit and Eclipse Fits}\label{transit_funct}  

We calculate our eclipse curve using the equations from \citet{mand02} assuming a circular orbit for the planet, as indicated by both radial velocity data \citep{bouchy05,triaud09,boisse09} and secondary eclipse photometry \citep{agol10,dewit12}.  We fix the planet's period to the value from \citet{knutson09a}, and take our limb-darkening coefficients in each bandpass from \citet{sing10}\footnote{Available at $www.astro.ex.ac.uk/people/sing$} assuming an \texttt{ATLAS} stellar atmosphere model with $T_\mathrm{eff}=5100$~K, $\log(g)=5.0$, and [Fe/H]=0 \citep{bouchy05}.  Our transit light curve has four free parameters, including the planet-star radius ratio $R_p/R_{\star}$, the orbital inclination $i$, the ratio of the orbital semi-major axis to the stellar radius $a/R_{\star}$, and the transit time.  For the secondary eclipse we approximate the planet as a uniform disk \citep[for more on this see][]{agol10} where the depth is measured relative to the average of the phase curve function over the duration of the eclipse.  We account for the change in the planet's brightness during the secondary eclipse by rescaling the amplitudes of ingress and egress to match the values of the phase curve at the start and end of the eclipse, respectively.  Our secondary eclipse light curves use the same values of $R_p/R_{\star}$, $i$, and $a/R_{\star}$ as the transit, and we allow the depths and times of each secondary eclipse to vary individually.  This gives us a total of eight free parameters in our fits to describe the transit and two secondary eclipses.

\subsection{Phase Curve Functions}\label{phase_curve_funct}

We calculate our phase curve functions as the sum of a series of sine and cosine terms with periods equal to the planet's orbital period and its even harmonics as described in \citet{cowan08}.  This approach produces phase curves with shapes that are indistinguishable from those of the ``orange-slice" models we used previously in \citet{knutson07,knutson09a}, but with the advantage that they result in more physically motivated smoothly varying brightness distributions instead of discrete segments of constant brightness on the surface of the planet.  We tried fits including sine and cosine terms with periods equal to the planet orbit $P$, as well as higher-order harmonics of $P/2$, $P/4$, and $P/8$.  As a check we also tried fitting the odd harmonics $P/3$ and $P/6$ to the data; as discussed in \citet{cowan08}, these harmonics cannot be produced by a static planetary brightness map, and their amplitudes therefore provide a useful test for whether or not we are over-fitting the data.  We find that the odd harmonic terms have fitted amplitudes comparable to those of the $P/4$ and $P/8$ terms, and therefore elect to limit our fits to the $P$ and $P/2$ harmonics:

\begin{align}\label{eq4}
F(\lambda) & = & 1+c_1\cos(2\pi t/P)+c_2\sin(2\pi t/P)\nonumber \\
& & +c_3\cos(4\pi t/P)+c_4\sin(4\pi t/P)
\end{align}

where $c_1-c_4$ are free parameters in the fits and $t$ is the time in days from the predicted center of transit.  

\begin{figure}
\epsscale{1.25}
\plotone{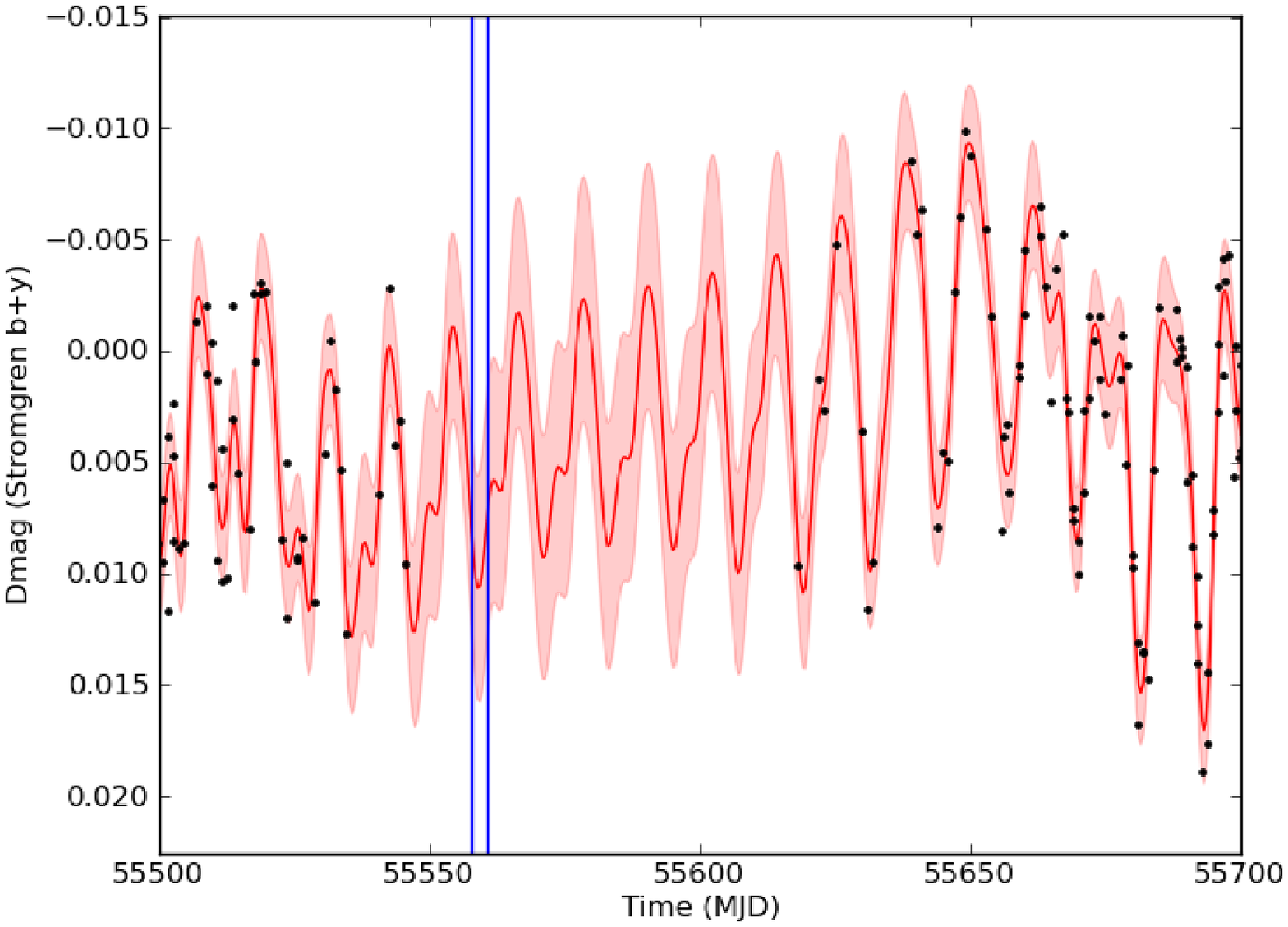}
\caption{Visible-light flux variations measured for HD~189733 in the averaged Str\"omgren \emph{b} and {y} filters (black circles) with the APT spanning the epoch of our 3.6~\micron~phase curve observation (vertical blue lines).  The solid red line indicates the best-fit spot solution with $1\sigma$ uncertainties delineated by the shaded pink region, which we use to extrapolate the likely stellar flux variations during our phase curve observation.}
\label{ch1_spots}
\end{figure}

\begin{figure}
\epsscale{1.25}
\plotone{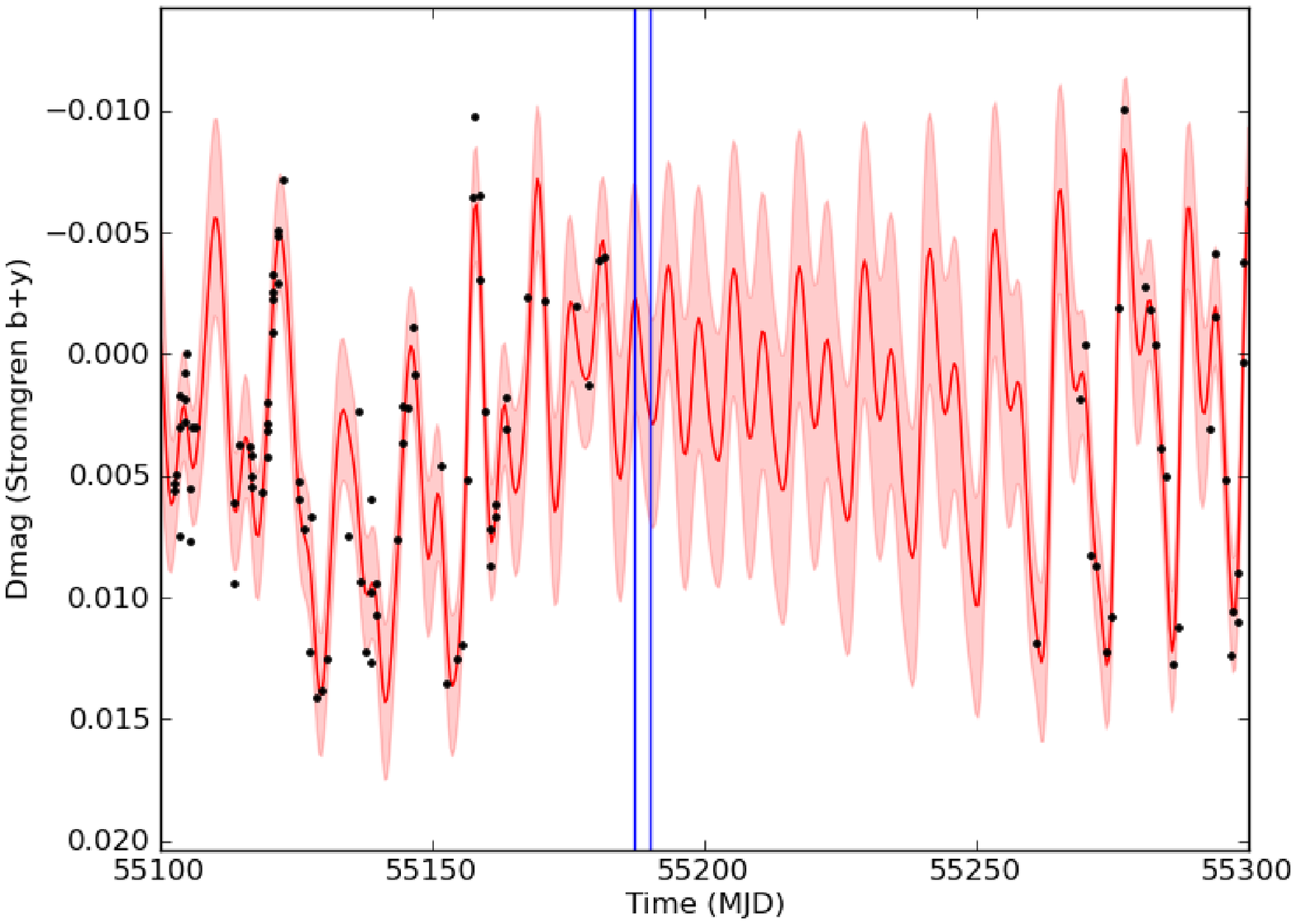}
\caption{Visible-light flux variations measured for HD~189733 in the averaged Str\"omgren \emph{b} and {y} filters (black circles) with the APT spanning the epoch of our 4.5~\micron~phase curve observation (vertical blue lines).  The solid red line indicates the best-fit spot solution, with $1\sigma$ uncertainties delineated by the shaded pink region, which we use to extrapolate the likely stellar flux variations during our phase curve observation.}
\label{ch2_spots}
\end{figure}

\subsection{Stellar Variability}\label{star_spots}

HD~189733b is an active K0 star with a Ca II H \& K emission line strength of \logr$=-4.5$ \citep{bouchy05,knutson10}, which has been observed to vary by as much as $\pm1.5$\% at visible wavelengths with a rotation period of 11.95 days \citep{henry08,knutson09a,lanza11}.  Although the amplitudes of these variations are reduced in the infrared, they are still comparable to the planet's phase curve and must be accounted for in our estimates of the phase curve amplitude \citep{knutson09a} and wavelength-dependent transit depth \citep{pont08,desert09,desert11,sing11}.  We characterize the nature of these trends using ground-based monitoring data for HD~189733b obtained with the Tennessee State University T10 0.8~m automated photometric telescope (APT) at Fairborn Observatory as part of a long-term monitoring program extend over a period of six years in total \citep{henry08}.  The data were obtained in the Str\"omgren \emph{b} and {y} filters, nodding between HD 189733 and three comparison stars of comparable or greater brightness as described in \citet{henry08}.  We present the averaged photometry spanning the epochs of the 3.6 and 4.5~\micron~\emph{Spitzer} data in Fig. \ref{ch1_spots} and \ref{ch2_spots}.  

\begin{figure}
\epsscale{1.1}
\plotone{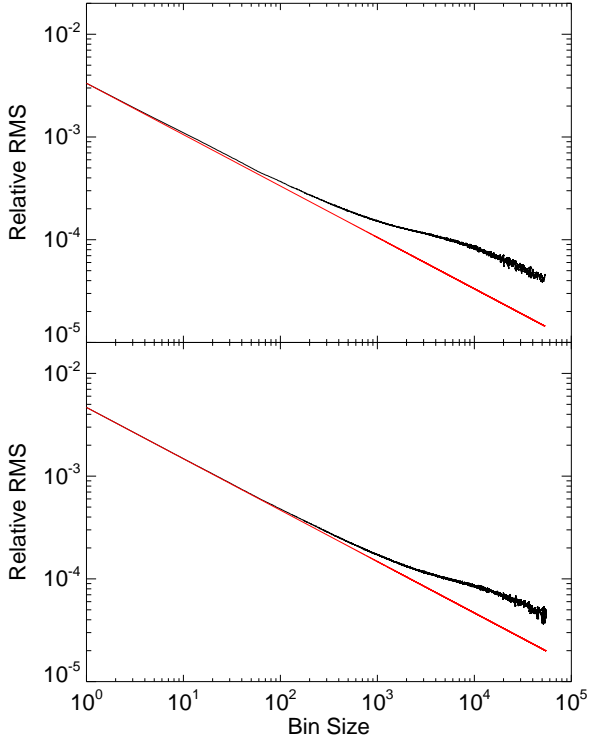}
\caption{Standard deviation of residuals vs. bin size for the 3.6~\micron~(top panel) and 4.5~\micron~(bottom panel) data after correcting for intrapixel sensitivity variations and dividing out the best-fit solutions for the transit, secondary eclipses, phase variations, and stellar activity.  Bin sizes of $10^3$ and $10^4$ correspond to time intervals of approximately 2 minutes and 20 minutes, respectively.}
\label{rootn}
\end{figure}

Because HD~189733 was not observable from the ground during the dates of the \emph{Spitzer} observations, we must extrapolate from ground-based measurements obtained several weeks prior and following our observations in order to infer the behavior of the star during this period.  We fit the data shown in Fig. \ref{ch1_spots} and \ref{ch2_spots} with a spot model for the star following the method described in \citet{aigrain11}, with uncertainties calculated using a Gaussian processes method described in \citet{gibson11}.  Based on these models, we find that the star varied in brightness by approximately $0.4\%-1.8$\% (3.6~\micron) and $0.3\%-1.4\%$ (4.5~\micron) in the averaged Str\"omgren \emph{b} and {y} filters during our phase curve observations, with a local minimum in the 3.6~\micron~data and a linearly decreasing flux in the 4.5~\micron~data.  If we rescale these variations to the wavelengths of our \emph{Spitzer} observations assuming a spot temperature of 4250~K \citep[e.g.,][]{pont08}, we find corresponding amplitudes of $0.1\%-0.4\%$ at 3.6~\micron~and $0.1\%-0.5\%$ at 4.5~\micron.  We account for the effects of the spots in our data by including a quadratic function of time in our 3.6~\micron~fit and a linear function of time in our 4.5~\micron~fit:  

\begin{equation}\label{eq5}
F(t) = d_1t+d_2t^2
\end{equation}
where $t$ is defined as the time from the predicted center of transit, $d_1$ and $d_2$ are free parameters in the 3.6~\micron~fit, and we set $d_2$ equal to zero in the 4.5~\micron~fit.  As a test of our ability to distinguish between spot variability and the planet's phase curve in these bands, we inserted linear functions of time with slopes ranging between $\pm0.5$\% in both data sets.  We find that we are able to retrieve the correct slope for our inserted trend in all cases, although our nominal phase curve solutions displayed some degeneracies with these trends for amplitudes larger than $0.3\%$.  Including a quadratic function of time in our fits to the the 4.5~\micron~data produces a negligible improvement in the Bayesian Information Criterion (BIC; 1,651,608 for the linear fit vs. 1,651,584 for the quadratic fit)\footnote{Defined as BIC$=\chi^2+k\ln(n)$, where k is the degrees of freedom and n is the total number of points in the fit \citep[e.g.][]{stevenson10}.}.  As an additional test, we generated 100 random samples from the Gaussian process model trained on the visible-light photometry.  We then fit the predicted stellar trend from each sample during the epoch of our 4.5~\micron~phase curve observation with both a linear and a quadratic function of time.  We find that the addition of a quadratic term produces a significant ($\Delta BIC>5$) improvement less than 30\% of the time, in good agreement with our previous test.  We therefore limit our 4.5~\micron~fits to a linear function of time in the analysis described below and find that the star decreased in brightness by $0.180\%\pm0.022\%$ over the course of our observations (see Table \ref{global_param}), in good agreement with our prediction of $0.1\%-0.5\%$ from the visible-light photometry.

In contrast to this result, we find that at 3.6~\micron~the addition of a quadratic term to our stellar spot function results in a significantly improved BIC (1,604,852 for the linear fit vs. 1,604,538 for the quadratic fit).  This is consistent with the results from our sampling of the Gaussian process model, which indicate that the quadratic term provides a better description for the predicted trend during our 3.6~\micron~phase curve observations in more than 70\% of cases.  However, we find that the best-fit quadratic solutions from our fits to the visible-light and 3.6~\micron~photometry are mutually inconsistent; we strongly prefer a negative quadratic term and a local flux maximum in our 3.6~\micron~fits, whereas the visible-light photometry indicates a positive quadratic term and a local flux minimum in 83\% of our solutions.  Upon closer examination we noticed that our best-fit quadratic function of time from the 3.6~\micron~photometry produced a local maximum just before the secondary eclipse, around the same time where we would expect to see a maximum in our best-fit phase curve.  We therefore conclude that the quadratic fit is degenerate with our phase curve function in this bandpass and opt to limit the quadratic coefficient in our subsequent fits to positive values (i.e., a linear function or a local flux minimum) in order to avoid this degeneracy and ensure good agreement with our visible-light spot models.   Our best-fit solution in this band is very close to linear and indicates that the star increased in brightness by $0.262\%\pm0.016\%$ during our observations (see Table \ref{global_param}, in good agreement with our prediction of $0.1\%-0.4\%$ from the visible-light photometry.

\begin{figure}
\epsscale{1.2}
\plotone{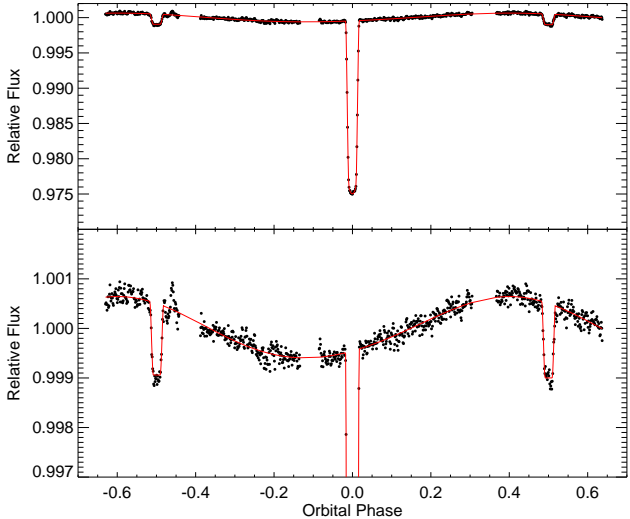}
\caption{Final 3.6~\micron~photometry after correcting for intrapixel sensitivity variations and stellar activity (black filled circles), binned into four minute intervals.  The best-fit phase, transit, and eclipse curves are overplotted as a red line.  The lower panel shows the same data as the upper panel, but with an expanded y axis for a better view of the phase curve.}
\label{ch1_phase}
\end{figure}

\section{Results}\label{results}

\begin{figure}
\epsscale{1.2}
\plotone{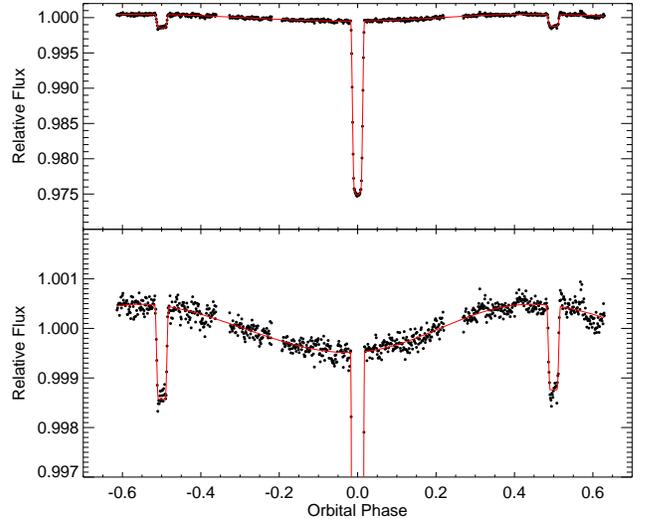}
\caption{Final 4.5~\micron~photometry after correcting for intrapixel sensitivity variations and stellar activity (black filled circles); see Fig. \ref{ch1_phase} for a complete description.}
\label{ch2_phase}
\end{figure}

We fit the trimmed data and calculate uncertainties simultaneously for the transit, secondary eclipses, phase curve, stellar activity, and intrapixel sensitivity corrections using a Markov Chain Monte Carlo (MCMC) method \citep[see, for example][]{ford05,winn07b} with a total of $10^5$ steps and either 16 (3.6~\micron) or 15 (4.5~\micron) free parameters.  These parameters include: the four phase function coefficients $c_1-c_4$, $a/R_{\star}$, $i$, $R_P/R_{\star}$, transit time, two secondary eclipse depths, two secondary eclipse times, either a linear (4.5~\micron) or quadratic (3.6~\micron) function of time to account for stellar variability, and two noise parameters discussed below.  We plot the normalized time series after the best-fit intrapixel sensitivity variations and stellar trends have been removed in Fig. \ref{ch1_phase} and \ref{ch2_phase}.

We find that there is still some remaining time-correlated noise in the residuals from our best-fit solution (Fig. \ref{rootn});  we account for this time correlation by implementing a wavelet-based MCMC fit as described in \citet{carter09} and compare the results of this fit to the standard $\chi^2$-based methods described in \citet{ford05} and \citet{winn07b}.  For the standard MCMC fit, which assumes that the uncertainties on individual points are Gaussian and time-independent, we set the per-point uncertainty equal to the value needed to produce a reduced $\chi^2$~equal to one for the best-fit solution (0.333\% at 3.6~\micron~and 0.466\% at 4.5~\micron).  In our wavelet MCMC we maximize the likelihood function instead of minimizing $\chi^2$, which allows us to fit for the white ($\sigma_w$) and red ($\sigma_r$) noise contributions to the per-point uncertainties.  The wavelet transform routine requires our data to have a length equal to a power of two, which we achieve by subdividing each phase curve into thirteen segments of equal length and zero-padding by either 6\% (3.6~\micron) or 3\% (4.5~\micron).  We prefer zero-padding over trimming as it allows us to include all available data in our fits, and we find that this approach has a negligible effect on our best-fit noise parameters.  

\begin{deluxetable*}{lrrrrcrrrrr}
\tabletypesize{\scriptsize}
\tablecaption{Global Fit Parameters \label{global_param}}
\tablewidth{0pt}
\tablehead{
\colhead{Parameter} & \colhead{\phantom{000000000} 3.6~\micron} & \colhead{\phantom{0000000} 4.5~\micron} }
\startdata
\emph{Transit Parameters} &&\\
$R_p/R_{\star}$\tablenotemark{a} & $0.15511\pm0.00020$ & $0.15580\pm0.00019$\\ 
$i$(\degr) & $85.671\pm0.034$ & $85.735\pm0.036$ \\
$a/R_{\star}$ & $8.858\pm0.031$ & $8.902\pm0.032$ \\
$T_c$ (BJD)\tablenotemark{b} & $2455559.554550\pm0.000035$ & $2455189.052491\pm0.000032$\\
&&\\
\emph{Secondary Eclipse Parameters} &&\\
1$^{\text{st}}$ Eclipse Depth & $0.1440\% \pm0.0053\%$ & $0.1889\% \pm0.0056\%$ \\
$T_c$ (BJD)\tablenotemark{b} & $2455558.44753\pm0.00081$ & $2455187.94447\pm0.00040$\\
2$^{\text{nd}}$ Eclipse Depth & $0.1500\%\pm0.0061\%$ & $0.1696\%\pm0.0053\%$ \\
$T_c$ (BJD)\tablenotemark{b} & $2455560.66515\pm0.00046$ & $2455190.16158\pm0.00048$\\
&&\\
\emph{Phase Curve Parameters (Eq. \ref{eq4}) } &&\\
 $c_1$ ($\cos(2\pi t/P)$ term)& $-0.0479\% \pm0.0032\%$ & $-0.0468\% \pm0.0044\%$ \\
 $c_2$ ($\sin(2\pi t/P)$ term) & $-0.0389\% \pm0.0028\%$ & $0.0122\% \pm0.0027\%$ \\
 $c_3$ ($\cos(4\pi t/P)$ term) & $0.0025\% \pm0.0021\%$ & $-0.0011\% \pm0.0019\%$ \\
 $c_4$ ($\sin(4\pi t/P)$ term)  &  $-0.0020\% \pm0.0014\%$ & $-0.0020\% \pm0.0022\%$ \\
Amplitude & $0.1242\% \pm 0.0061\%$ & $0.0982\pm0.0089\%$ \\
Minimum Flux Offset\tablenotemark{c} [h] & $-6.43\pm0.82$ & $-1.37\pm1.00$ \\
Maximum Flux Offset\tablenotemark{c} [h] & $-5.29\pm0.59$ & $-2.98\pm0.82$ \\
&&\\
\emph{Stellar Flux Variation (Eq. \ref{eq5}) } &&\\
Linear term $d_1$ & $0.000932\pm0.000057$ & $-0.000652\pm0.000079$ \\
Quadratic term $d_2$ & $<0.0000041$ & 0 (fixed) \\
&&\\
\emph{Noise Parameters} &&\\
$\sigma_w$ & $0.0032608\pm0.0000027$ & $0.0046009\pm0.0000037$ \\
$\sigma_{r}^2$\tablenotemark{e} & $0.000581\pm0.000043$ & $0.000283\pm0.000037$ \\
\enddata
\tablenotetext{a}{These values do not include a correction for stellar variability; see \S\ref{depth_discussion} for the corrected values.}
\tablenotetext{b}{We list all times in BJD\_UTC for consistency with other studies; to convert to BJD\_TT, simply add 66.184~s.}
\tablenotetext{c}{Offsets are measured relative to the center of transit time for the minimum flux and the center of secondary eclipse time for the maximum flux.  A negative offset indicates that the maximum or minimum occurs earlier than predicted, corresponding to an eastward offset in the location of the hot or cold region in the planet's atmosphere.}
\tablenotetext{d}{This parameter is limited to values greater than or equal to zero in our fits; see \S\ref{star_spots} for more information.  We find that our best-fit distribution peaks at zero and we therefore report the $1\sigma$ upper limit here.}
\tablenotetext{e}{Unlike $\sigma_w$, the red noise parameter is not equal to the variance of the red noise component of the time series; see \citet{carter09} for a more detailed explanation.}
\end{deluxetable*}

We initialize each chain at a position determined by randomly perturbing the best-fit parameters from a Levenberg-Marquardt minimization.  After running the chain, we search for the point where the $\chi^2$ value first falls below the median of all the $\chi^2$~values in the chain (i.e. where the code had first found the optimal fit), and discard all steps up to that point.  For the wavelet MCMC we perform a similar trim where the likelihood first rises above the median value.  We set our best-fit parameters equal to the median of each distribution and calculate the corresponding uncertainties as the symmetric range about the median containing 68\%~of the points in the distribution.  The distribution of values was very close to symmetric for all parameters, and there did not appear to be any detectible correlations between variables aside from $i$ and $a/R_{\star}$.  Our wavelet analysis produces modestly higher uncertainties for most parameters, with the greatest increases in the planet-star radius ratio and secondary eclipse depth errors, which were a factor of two larger than in the standard Gaussian $\chi^2$ fits.  We list the best-fit parameters and corresponding wavelet-based uncertainties in Table \ref{global_param}.  

For both fits the standard deviation of our best-fit residuals is a factor of 1.12 higher than the predicted photon noise limit at 3.6~\micron, and 1.13 times higher than this limit at 4.5~\micron.  If we calculate the standard deviation of the red noise vector in our best-fit wavelet solution, we find that the red noise contributes 0.0162\% (3.6~\micron) and 0.0017\% (4.5~\micron) of the total scatter in the unbinned residuals.   We also estimate the red noise on time scales relevant for the eclipses (i.e., approximately one hour) by comparing the standard deviation of our best-fit residuals to the theoretical Poisson limit for a bin size of 27,000 points in Fig. \ref{rootn}.  We find that red noise contributes a scatter of approximately 0.003\% in relative flux in both bandpasses, corresponding to a factor of 2.5 (3.6~\micron) or 2.1 (4.5~\micron) increase in the total noise for our one hour bins.  This is consistent with the results of our wavelet analysis, which finds that the uncertainties in the measured secondary eclipse depths in both channels are approximately double those predicted assuming Gaussian noise in the standard MCMC fits.

\begin{figure*}
\epsscale{1.2}
\plotone{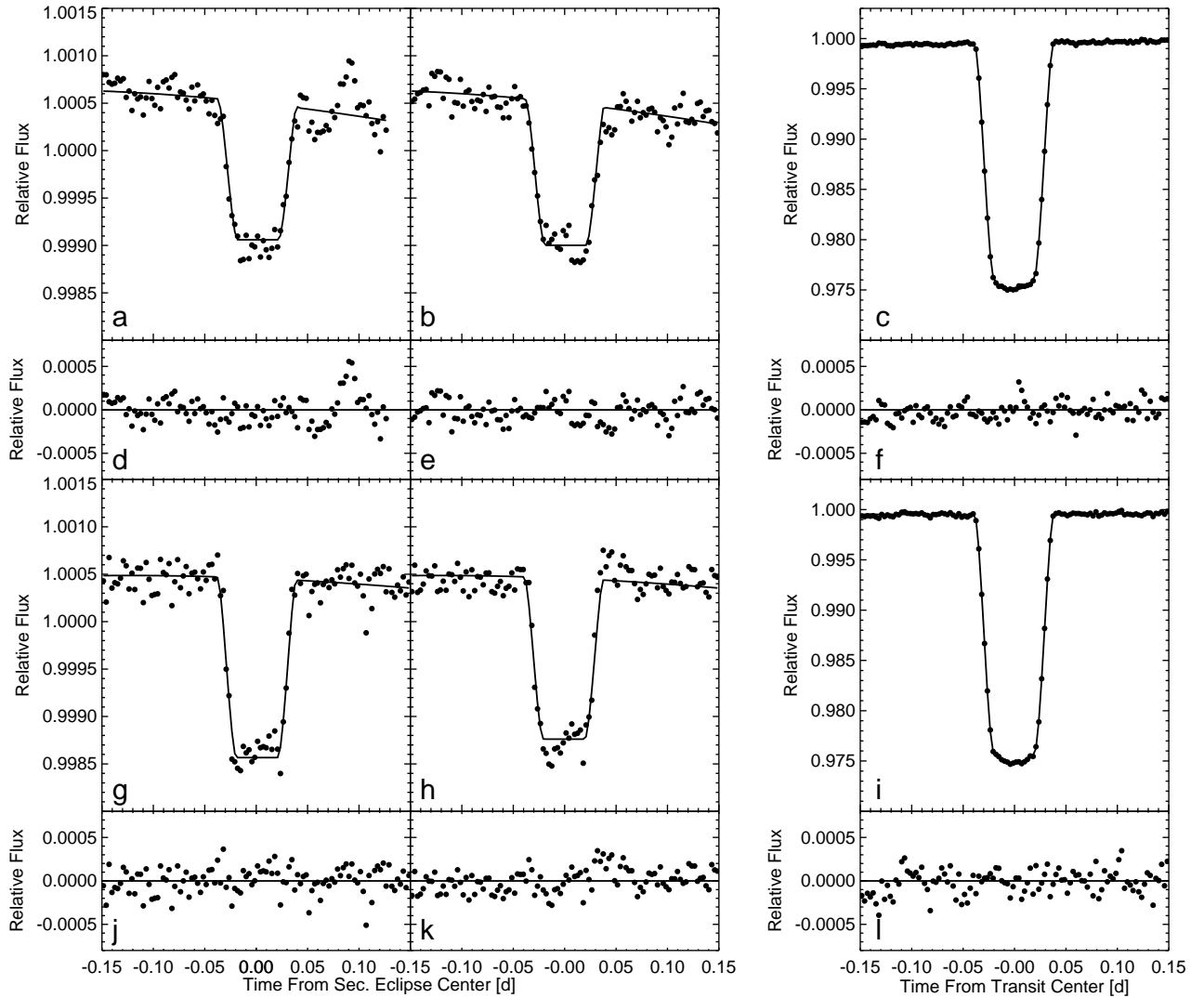}
\caption{Best-fit transit and secondary eclipse light curves for the 3.6~\micron~(a$-$f) and 4.5~\micron~(g$-$l) bands (black filled circles).  The data have been binned in four minute intervals.  The first secondary eclipses are shown on the left (a and g), the second secondary eclipses are shown in the middle (b and h), and the two transits are shown on the right (c and i).  Residuals from the best-fit solution (black line, a$-$c and g$-$i) are shown immediately beneath each event in panels d$-$f and j$-$l.}
\label{combined_eclipses}
\end{figure*}

\section{Discussion}

We present the results of our transit and secondary eclipse fits in the sections below, focusing on the measured depths and times of eclipse.  We find an average orbital inclination of $85.701\degr \pm0.025\degr$, $a/R_{\star}=8.879\pm0.022$, and an impact parameter $b=0.6656\pm0.0042$, consistent with previous estimates from \citet{pont07}, \citet{sing09}, \citet{agol10}, and \citet{desert09,desert11}.  We note that our errors are smaller than those reported by \citet{desert11} for their 3.6~\micron~transit observation; this is likely due to the availability of a longer baseline and correspondingly tighter constraints on the correction for the intrapixel sensitivity variations in our observations. 

\subsection{Orbital Ephemeris}\label{ephemeris_discussion}

\begin{figure}
\epsscale{1.2}
\plotone{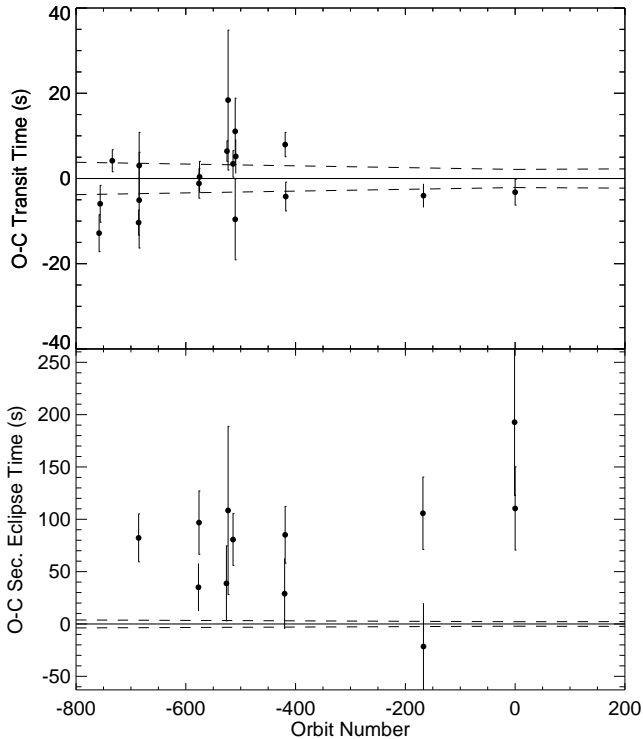}
\caption{Observed minus calculated transit (top panel) and secondary eclipse (bottom panel) times from the published literature.  The two most recent transit measurements and four most recent secondary eclipse observations are derived from the observations presented in this paper; the timing offsets in earlier transits are likely the result of occulted star spots.  We calculate predicted secondary eclipse times assuming an offset of half an orbit from the transit center and use the ephemeris presented in \S\ref{ephemeris_discussion}.  Dashed lines indicate the $1\sigma$ uncertainties in the predicted transit and secondary eclipse times.}
\label{ephemeris}
\end{figure}

We show the best-fit transit and secondary eclipse solutions in Fig. \ref{combined_eclipses}; we see no evidence for excess scatter in the transit residuals, suggesting that our assumptions for the shape of the planet (i.e., oblate vs. circular) and limb-darkening coefficients are correct to within our uncertainties.  We calculate an updated ephemeris for the system using the transit times given in Table \ref{global_param} and previously published values from \emph{Spitzer} \citep{knutson09a,desert09,desert11,agol10} and \emph{Hubble} \citep{pont07}.  We define this ephemeris as:

\begin{align}\label{eq3}
T_c(n)= T_c(0)+n\times P
\end{align}
where $T_c$ is the predicted transit time as a function of the number of transits elapsed since $T_c(0)$ and $P$ is the orbital period.  Our new observations extend the previous baseline by more than two years, and as a result we derive a more precise estimate of the planet's orbital ephemeris: $T_c(0)=2455559.554587\pm0.000025$~BJD and $P=2.218575143\pm0.000000046$~days ($\pm4$~ms), where $T_c(0)$ and $P$ are determined from a linear fit to our measured transit times and the previously published values listed above.  As shown in Fig. \ref{ephemeris}, our new transit times are consistent with a constant ephemeris in all cases.  

We fit a second, independent ephemeris to the secondary eclipse times from  \citet{charbonneau08}, \citet{knutson09a}, and \citet{agol10}, as well as this paper, and find a best-fit value of: $T_0=2455560.66451\pm0.00025$~BJD and $P=2.218574560\pm0.000000454$~days.  This period differs from the best-fit transit period by $1.3\sigma$.

We find that the secondary eclipse times occur systematically later than predicted, with an average delay of $49\pm9$~s after accounting for the 31~s. light travel time delay \citep{agol10}.  The additional delay is consistent with the time offset predicted for a planet with a non-uniform surface brightness, which alters the shape of ingress and egress and the corresponding best-fit eclipse time estimate \citep{williams06}.  In \citet{agol10} we found that the eastward offset of the hot spot on the planet's day side causes the best-fit 8~\micron~eclipse time to shift approximately 30~s later than the true value.  This result was recently confirmed by \citet{majeau12} and \citep{dewit12}, who used a variety of methods to create 2D maps of HD~189733b's 8~\micron~dayside brightness distribution based on the shape of its 8~\micron~phase curve and secondary eclipse ingress and egress.  If we examine our new secondary eclipse times separately we find average offsets of $99\pm35$~s at 3.6~\micron~and $22\pm27$~s at 4.5~\micron~after accounting for the light travel time.  Although these estimates are consistent with the 8~\micron~offsets, they are not precise enough to provide independent constraints on the location of the dayside hot spot in the 3.6 and 4.5~\micron~bands. 

\subsection{Transit and Secondary Eclipse Depths}\label{depth_discussion}

We obtain planet-star radius ratios of $0.15511\pm0.00020$ at 3.6~\micron~and $0.15580\pm0.00019$ at 4.5~\micron, which we compare to previously published values of $0.1545\pm0.0003$ at 3.6~\micron~and $0.1557\pm0.0003$ at 4.5~\micron~from \citet{desert09}, and $0.15566^{0.00011}_{-0.00024}$ at 3.6~\micron~from \citet{desert11}.  Although our results are consistent with these previous measurements at the $1.7\sigma$ (3.6~\micron) and $0.3\sigma$ (4.5~\micron) level, we note that variations in the stellar flux between epochs can affect the measured transit depths, and must be accounted for if we wish to compare these two sets of measurements.  In this case we are most interested in the relative difference between our measured 3.6 and 4.5~\micron~radius ratios, which can be used to search for the presence of absorption features in HD~189733b's transmission spectrum.  \citet{desert11} find that their 4.5~\micron~radius ratio is $0.0012\pm0.0004$ smaller than at 3.6~\micron~and suggest that this may be consistent with absorption from carbon monoxide, although they do not correct explicitly for changes in the stellar flux over the approximately one year interval between their two measurements.  

We account for the effect of varying stellar flux in our data using the ground-based monitoring data described in \S\ref{star_spots}.  We find that the star is 0.9\%$\pm0.6$\% fainter in the Str\"omgren $(b+y)/2$ band during our 3.6~\micron~transit as compared to our 4.5~\micron~transit.  If we assume a spot temperature of $4250$~K as estimated by \citet{pont08}, we find that the corresponding flux change in the 3.6 and 4.5~\micron~bands should be approximately 28\% of the variation in the Str\"omgren $(b+y)/2$ band, or 0.3\%$\pm0.2$\%.  Correcting the 3.6~\micron~transit depth to the equivalent stellar flux at the time of the 4.5~\micron~transit observation yields a planet-star radius ratio of $0.15465\pm0.00037$, $0.00115\pm0.00042$ ($2.7\sigma$) smaller than in the 4.5~\micron~band.  This result is consistent with the planet-star radius ratio difference of $0.0012\pm0.0004$ reported by \citet{desert09}, who also find a deeper transit at 4.5~\micron~as compared to 3.6~\micron.  We note that although our measurements of the planet-star radius ratios are more precise than those reported in this paper, we are ultimately limited by the uncertainties in our estimates of the stellar flux and its corresponding effect on the measured transit depths at the epochs of our two measurements.

We also present new estimates of the secondary eclipse depth in the 3.6 and 4.5~\micron~bands.  We see no convincing evidence for variability in the planet's flux over the planet's orbit; our two 3.6~\micron~depths differ by $0.8\sigma$, and our 4.5~\micron~depths differ by $2.5\sigma$.  We estimate an average secondary eclipse depth of $0.1466\%\pm0.0040$\% at 3.6~\micron~and $0.1787\%\pm0.0038$\% at 4.5~\micron.  We utilize a \texttt{PHOENIX} stellar atmosphere model for the star and calculate corresponding brightness temperatures of $1328\pm11$~K and $1192\pm9$~K, respectively.  These values are shallower than our previous estimates of $0.256\pm0.014$\% at 3.6~\micron~and $0.214\pm0.020$\% at 4.5~\micron~from \citet{charbonneau08}, although the 4.5~\micron~depths are consistent at the $1.7\sigma$~level.  We note that the error estimates for our new secondary eclipse depths are a factor of $4-5$ smaller than these previous values; this is mainly due to the fact that the older observations cycled between all four \emph{Spitzer} IRAC bands in order to obtain multi-wavelength coverage during a single secondary eclipse event, leading to an effective cadence one-fifth that of our new staring-mode observations.  Although our new 3.6~\micron~secondary eclipse depth differs from our older estimate by $7.5\sigma$, the older observation had much larger intrapixel sensitivity variations that were partially degenerate with the secondary eclipse depth in this band.  Because our error estimates in the 2008 paper were based on conventional MCMC methods, which assume that the noise is Gaussian, it is likely that the uncertainties on the older eclipse depth were significantly underestimated.  Although we considered re-analyzing these older data with our new methods, we concluded that the lower cadence and larger pointing jitter in these observations meant that we were unlikely to get useful constraints on the secondary eclipse depth as compared to our new observations.

\subsubsection{Comparison to 1D Atmosphere Models}\label{1D_models}

\begin{figure}
\epsscale{1.2}
\plotone{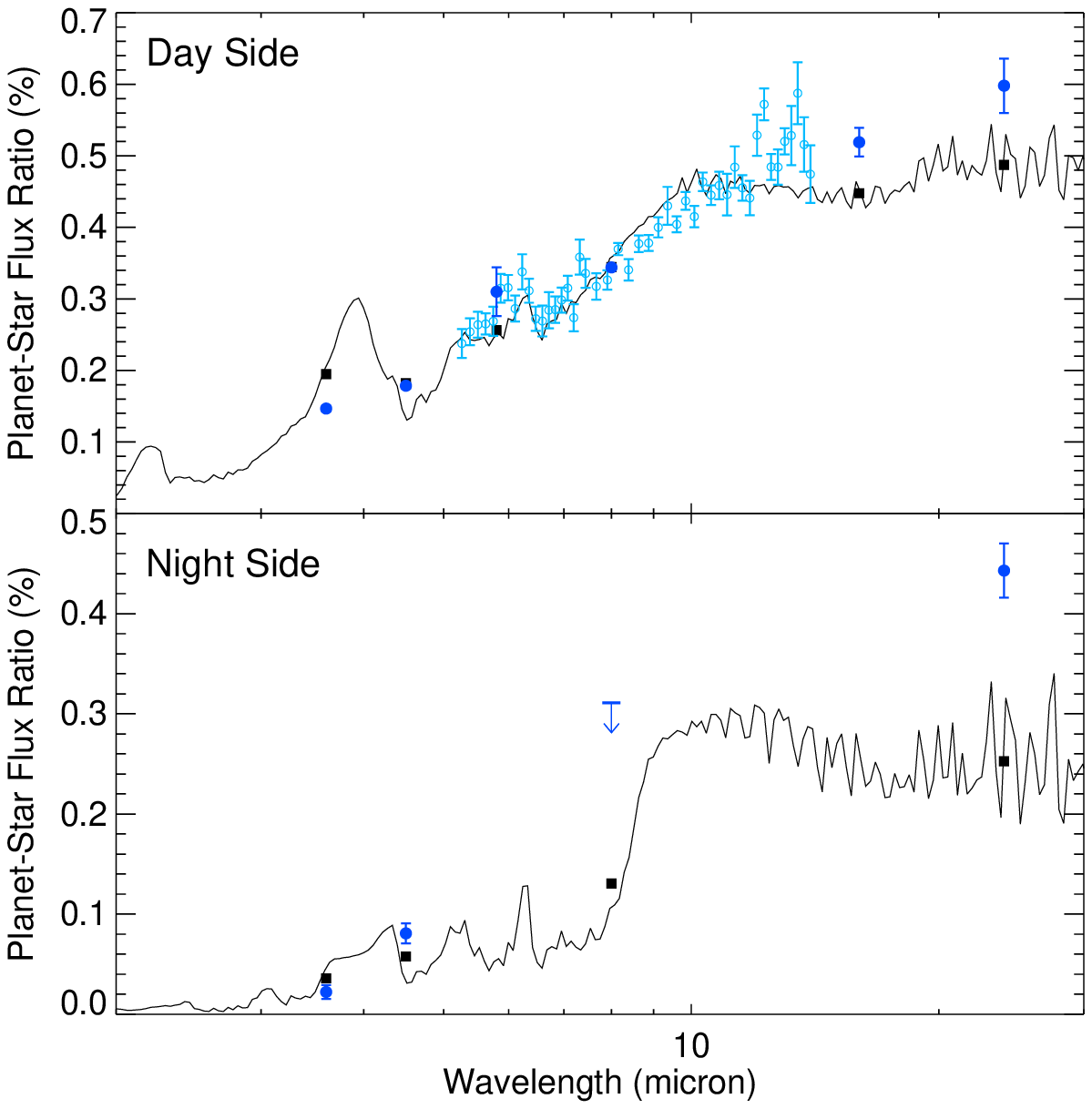}
\caption{Upper panel: measured broadband secondary eclipse depths (filled dark blue circles) at 3.6 and 4.5~\micron~(this paper), 5.6, 16, and 24~\micron~\citep{charbonneau08}, and 8~\micron~\citep{agol10}, and an IRS spectrum (open light blue circles, Grillmair et al. 2008) compared to a 1D atmosphere model for HD 189733b's day side (black line) following \citet{burrows08} with parameterized recirculation $P_n=0.15$ at depth and a high-altitude absorber with an opacity equal to $\kappa_e=0.035$ cm$^{-1}$.  Black filled squares show the predicted planet-star flux ratios in the same bands as the \emph{Spitzer} observations.  The lower panel shows the measured minimum fluxes for HD 189733 (filled dark blue circles) compared to the same 1D atmosphere model for the planet's night side; the temperature in this region is set by the amount of energy transported from the day side.}
\label{burrows}
\end{figure}

We compare our new secondary eclipse depths to the predictions of one-dimensional radiative transfer models for HD 189733b's dayside emission as described in \citet{burrows08}.  These models assume a solar metallicity atmosphere with the chemistry in local thermal equilibrium (LTE), an interpolated \texttt{ATLAS} model for the stellar spectrum, and allow for the parameterized redistribution of heat from the dayside to the night side using a heat sink between $0.01-0.1$~bars.  They also include an additional grey absorber at low pressures, which is required to reproduce the high-altitude temperature inversions observed in the atmospheres of a subset of hot Jupiters such as HD 209458b \citep[e.g.,][]{burrows07,knutson08,swain09,madhu10}.  Both the fractional amount of the incident flux $P_n$ redistributed to the night side and the opacity of the high-altitude absorber $\kappa_e$ in cm$^{-1}$ g are left as free parameters and optimized to provide the best match to the updated secondary eclipse depths at 3.6 and 4.5~\micron, together with previously published 5.8, 8.0, 16, and 24~\micron~eclipse depths from \citet{charbonneau08} and \citet{agol10} and the $5-12$~\micron~IRS spectrum from \citet{grillmair08}.  We obtain a reasonably good match to all measurements except the 3.6~\micron~dayside using a model with $P_n=0.15$ (i.e., 15\% redistribution, where 50\% redistribution corresponds to a perfectly well-mixed atmosphere) and $\kappa_e=0.035$ cm$^{-1}$ g, as shown in Fig. \ref{burrows}.  These values are identical to those of the nominal model presented in \citet{grillmair08}.  For this model, our measured dayside fluxes differ by [$-12.1\sigma$, $-0.9\sigma$, $+1.6\sigma$, $-0.1\sigma$, $+3.6\sigma$, $+2.9\sigma$] from their predicted values in the [3.6, 4.5, 5.8, 8.0, 16, 24]~\micron~bands, respectively.

Given that our models include an estimate of the relative day-night recirculation efficiency, it is worth considering whether or not a nightside model with the same assumptions as our day side model can accurately reproduce the measured night side fluxes at 3.6, 4.5, 8.0, and 24~\micron.  We take the longer-wavelength flux estimates from \citet{knutson09a} and estimate the 3.6 and 4.5~\micron~night side fluxes as the measured secondary eclipse depths minus the phase curve amplitude in each band; this gives planet-star flux ratios of $0.0224\%\pm0.0073\%$ and $0.0805\%\pm0.0097\%$ at 3.6 and 4.5~\micron, respectively.  If we use a \texttt{PHOENIX} stellar atmosphere model for the star, these fluxes correspond to brightness temperatures of $825\pm54$~K and $928\pm33$~K, respectively.  We compare these fluxes to the predictions of a solar metallicity LTE model with $P_n=0.15$; the model does not include any additional high-altitude absorber, but this is irrelevant for the night side as there is no incident flux at the top of the atmosphere.  We find that this default night side model provides a good match for the measured 3.6 and 4.5~\micron~nightside fluxes ($-1.9\sigma$ and $+2.3\sigma$ difference, respectively), although it underestimates the 24~\micron~flux by $7.1\sigma$.  We conclude that statistical surveys seeking to estimate the relative amount of day-night circulation based on fits to the dayside emission alone, such as those presented in \citet{cowan11}, should obtain reasonable estimates for this quantity despite the limited data available and the relative simplicity of the models used. 

\subsubsection{A Note on Stellar Atmosphere Models}

There are several potential sources of stellar atmosphere models for HD~189733 in the literature, including both interpolated and star-specific \texttt{ATLAS} models \citep{kurucz79,kurucz94,kurucz05}\footnote{Available at $http://kurucz.cfa.harvard.edu/$} and interpolated \texttt{PHOENIX} models \citep{hauschildt99}.  We compare the stellar spectra obtained from these three sources, and find that the star-specific \texttt{ATLAS} model over-predicts the strength of the CO absorption feature between $4-5$~\micron~as compared to our interpolated \texttt{ATLAS} and \texttt{PHOENIX} spectra.  If HD 189733's spectrum is best-described by this star-specific model it would shift the predicted planet-star flux ratio in the 4.5~\micron~\emph{Spitzer} band to larger values, at a level comparable to or larger than our uncertainties for this measurement.  For now we proceed using the interpolated grid models, which we find give consistent predictions regardless of which grid they are taken from. 

\subsection{Phase Curve Fits}\label{phase_discussion}

There are three basic quantities which we obtain from our phase curve observations: the amplitude of the observed variation, the location of  the flux maximum, and the location of the flux minimum.  Taken together, these quantities can be used to estimate the day-night brightness contrast as well as the positions of hot and cold regions in the atmosphere.  Because the opacities (and hence the pressure from which most of the detected flux is emitted) may vary from one band to the next, we do not necessarily expect to see the same features in all bandpasses.  This picture is further complicated by the fact that the wavelength-dependent opacity of the planet's atmosphere may vary as a function of longitude.  Our first task, therefore, is to consider whether or not our ensemble of phase curves all display the same qualitative features.

Focusing on the most basic feature first, we compare the amplitude of the phase variation in our four bands.  We find that the planet's minimum flux is $15\%\pm5\%$ of the maximum flux at 3.6~\micron, and $45\%\pm5\%$ at 4.5~\micron.  We can then convert this flux contrast to a brightness temperature contrast of $503\pm21$~K at 3.6~\micron~and $264\pm24$~K at 4.5~\micron.  The contrast in the 3.6~\micron~band is $6-7\sigma$ larger than in the 4.5~\micron~band or the corresponding value of $188\pm48$~K at 24~\micron~from \citet{knutson09a}.  If we assume that the composition of the planet's atmosphere is uniform (e.g., the opacity is constant), this would suggest that at longer wavelengths we are seeing deeper into the planet's atmosphere, where temperatures are expected to be more homogenous \citep[e.g.,][]{showman09,burrows10}.  This differs from the predictions of 1D solar metallicity atmosphere models assuming equilibrium chemistry  \citep[e.g.,][]{fortney08,burrows08,knutson09a}, which find that HD~189733b's 3.6~\micron~band should have the lowest average opacity and, therefore, probe deepest in the planet's atmosphere.  

We next compare the locations of the flux maxima in each band.  If the planet's atmosphere is in strict radiative equilibrium (i.e., no recirculation) we would expect the phase curve to reach its maximum at the center of the secondary eclipse, when we are viewing the substellar point face-on.  If, on the other hand, the atmospheric circulation is dominated by a strong west-to-east (i.e., superrotating) equatorial band of wind, we would expect to see the phase curve reach a maximum prior to the secondary eclipse.  We find that the maximum flux occurs $5.29\pm0.59$ hours before the center of the secondary eclipse at 3.6~\micron~and $2.98\pm0.82$ hours before the eclipse center at 4.5~\micron.  These two offsets differ from zero by $9.0$ and $3.6\sigma$, respectively, and from each other by $2.3\sigma$.

We next compare the locations of the minima in both light curves, which should be offset in the same direction as the maxima in the limit of a single equatorial band of wind.  We find that the minimum flux occurs $6.43\pm0.82$ hours before the transit center at 3.6~\micron~and $1.37\pm1.00$ hours before the transit center at 4.5~\micron.  These two measurements differ from zero by $7.8$ and $1.4\sigma$ respectively and are consistent with the presence of an eastward jet on the night side.  If this same eastward offset was present in our 8.0 and 24~\micron~phase curves it would have gone undetected, as our observations began shortly before the transit and spanned just half an orbit in both cases.  General circulation models \citep[e.g.,][]{showman09,burrows10,heng11} predict that in cases where a super-rotating eastward jet dominates heat transport the cold region on the planet's night side should display a larger shift than the hot region on the dayside; this results in maximum and minimum fluxes in the integrated light curves that are separated by slightly less than 180\degr.  We find that our data are consistent with a 180\degr~separation in both bands, but this is not surprising as our constraints are weak compared to the size of the predicted deviations. 

\subsubsection{The Location of the 8~\micron~Phase Curve Minimum}

\emph{Spitzer} data taken in the 8~\micron~band display a ramp-like behavior where higher-illumination pixels converge to a constant value within the first hour of observations while lower-illumination pixels show a continually increasing linear trend on the time scales of interest here.  This effect is generally attributed to charge-trapping in the array and is discussed in detail in \citet{knutson07,knutson09c} and \citet{agol10}, among others.  We must correct for this effect in order to retrieve the phase curve presented in \citet{knutson07}, which displays a flux minimum several hours after the transit (i.e., shifted in the opposite direction of the minima in our new 3.6 and 4.5~\micron~phase curves).  Although we initially considered this minimum to be reliably detected despite the uncertainties in our ramp correction, an independent analysis presented in \citet{agol10} used a different method to correct the ramp and found much larger uncertainties at early times.  

In this section we investigate the nature of the detector ramp in more detail, focusing on the question of whether or not the flux minimum in the 8~\micron~data might be either a direct result of the ramp or an artifact created by our ramp correction.  We utilize 8~\micron~preflash data obtained on UT 2009 Apr 23 as part of a program to observe CoRoT-7b \citep{fressin12}, which targeted a region in the Orion Nebular Cloud with bright, diffuse emission located at J2000 05:35:16 -05:23:24.  The purpose of these observations was to mitigate the ramp effect by filling the charge traps throughout the array immediately before slewing to the science target \citep[e.g.,][]{seager09,knutson09c}.  These data are particularly useful for characterizing the ramp effect because they allow us to bin pixels according to their median illumination level without worrying about added variability introduced by shifts in the telescope pointing.  For a bright, compact source such as a star, it is almost impossible to disentangle these two effects at the pixel level.

\begin{figure}
\epsscale{1.2}
\plotone{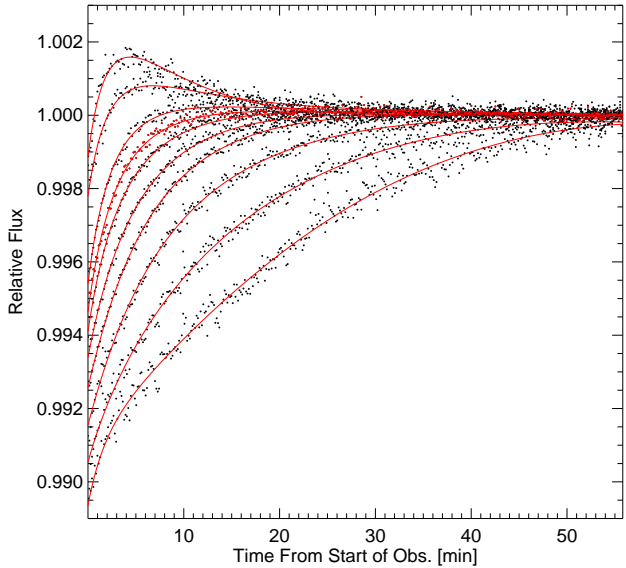}
\caption{Measured flux as a function of time in the 8~\micron~array for pixels with illumination levels ranging from $1000-2000$~MJy Sr$^{-1}$ (bottom curve) to $4000-45000$~MJy Sr$^{-1}$ (top curve), with best-fit fourth-order polynomial functions of time (red lines) overplotted for comparison.  Although the detector ramp typically resembles an asymptote that converges on time scales related to the median illumination level, the most highly illuminated pixels exhibit an additional over-shooting effect that could lead to residuals in the corrected light curves if not taken into account in our choice of functional form to describe the ramp.}
\label{preflash}
\end{figure} 

We divide the data from the preflash images into twelve bins, with illumination levels ranging from $1000-4500$~MJy Sr$^{-1}$.  Although these illumination levels are generally higher than the peak flux of approximately $1700$~MJy Sr$^{-1}$ reached in the central pixel of our 8~\micron~HD 189733b observations, we expect that lower illumination pixels will display qualitatively similar behaviors on longer time scales.  We plot the resulting light curves for each bin in Fig. \ref{preflash}.  Although the majority of pixels exhibit the typical asymptote associated with this effect, we note that the highest illumination pixels display an additional behavior which we term ``overshooting'', in which the flux increases past its equilibrium value and then gradually decreases until it reaches its nominal steady state.  It is possible that this effect is present in the lower-illumination light curves as well, but is hidden by the larger size of the initial ramp.  There is some evidence for this theory, as \citet{laughlin09} report an effect resembling overshooting in their 8~\micron~observations of the stellar binary component in the HD 80606 system.  In \citet{crossfield12}, we also find evidence for a similar effect in the 24~\micron~MIPS array. 

We note that exponential fits to the 8~\micron~ramp, such as those used in \citet{agol10}, do not take this effect into account.  Although we used a more general seventh-order polynomial function of time to describe the ramp in \citet{knutson07}, we only corrected a subset of the lower-illumination pixels inside our photometric aperture.  If the uncorrected higher-illumination pixels in our aperture displayed this behavior, or if it was inadequately described by our polynomial fits to the lower-illumination pixels, then we would expect to see a local minimum near the start of our observations.  Based on these data, we conclude that the flux minimum observed in our corrected data could reasonably be attributed to this overshooting effect, rather than the planet's phase curve.  For the purposes of comparing our 8~\micron~phase curve to other wavelengths, we adopt the convention of \citet{agol10} and trim the first part of the light curve where the ramp correction is largest and our conclusions correspondingly uncertain.  This trimming does not affect our estimate of the location of the flux maximum, but it does prevent us from determining the location of the flux minimum in these data.

\begin{figure}
\epsscale{1.2}
\plotone{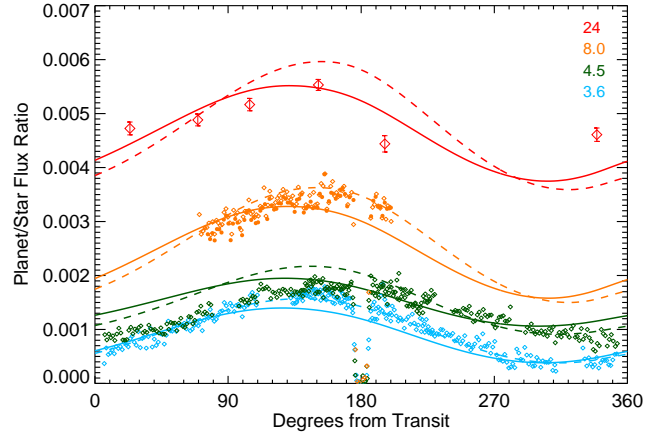}
\caption{Measured phase variations in the 3.6 (blue circles), 4.5 (green circles), 8.0 (orange open circles from Knutson et al. 2007, 2009a, orange filled circles from Agol et al. 2010), and 24~\micron~(red diamonds) \emph{Spitzer} bands after correcting for stellar flux variations and instrument effects.  Overplotted lines indicate the predictions of general circulation models for this planet in each band from \citet{showman09}, assuming either solar (solid) or $5\times$ solar (dashed) metallicity.}
\label{1x_5x_phase_curves}
\end{figure}  

\subsection{Comparison to General Circulation Models}

\begin{figure}
\epsscale{1.2}
\plotone{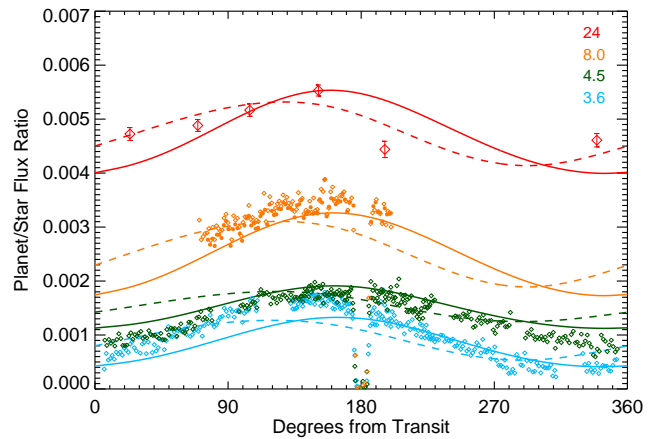}
\caption{Same data as in Fig. \ref{1x_5x_phase_curves}.  The overplotted lines indicate the predictions of general circulation models for this planet in each band from \citet{showman09}, assuming either a slowly rotating ($0.5\times$ orbital period; solid line) or rapidly rotating ($2\times$ orbital period; dashed line) planet.}
\label{slow_fast_phase_curves}
\end{figure}

In this section we combine our new 3.6 and 4.5~\micron~phase curves with our previous observations at 8.0 and 24~\micron~\citep{knutson09a} and compare these results to the predictions of 3D general circulation models from \citet{showman09}.  These models use the MITgcm to solve the primitive equations in three dimensions, and are coupled to a non-grey radiative transfer scheme.  Showman et al. considered four complementary cases for the planet, corresponding to: 1) a solar metallicity atmosphere and a 1:1 ratio for the planet's rotation and orbital periods, 2) a $5\times$ solar metallicity atmosphere and 1:1 rotation, 3) a solar metallicity atmosphere and a 1:2 rotation period (i.e., more slowly rotating than the tidally locked case), and 4) a solar metallicity atmosphere and a 2:1 rotation period (i.e., the rapid rotation case).  We calculate phase curves in the 3.6, 4.5, 8.0, and 24~\micron~\emph{Spitzer} bands for each model following the method described in \citet{fortney06}.  We use an interpolated \texttt{PHOENIX} model to calculate planet-star flux ratios and confirm that this model gives predictions consistent with those of the \texttt{ATLAS} model used with the 1D models in \S\ref{1D_models}.  We compare the resulting predictions to our measured phase curves in Fig. \ref{1x_5x_phase_curves} and \ref{slow_fast_phase_curves}; see Table \ref{phase_table} for a more quantitative comparison.

\begin{deluxetable*}{lrrrrcrrrrr}
\tabletypesize{\scriptsize}
\tablecaption{Phase Curve Comparison \label{phase_table}}
\tablewidth{0pt}
\tablehead{
\colhead{Source} & \colhead{3.6~\micron} & \colhead{4.5~\micron} & \colhead{8.0~\micron} & \colhead{24~\micron}}
\startdata
\emph{Maximum Planet-Star Flux Ratio\tablenotemark{a}} &&&\\
Measured\tablenotemark{b} & $0.1466\%\pm0.0040$\% & $0.1787\%\pm0.0038$\%& $0.344\%\pm0.004\%$ & $0.598\%\pm0.038\%$ \\ 
1x Solar Model & 0.1397\%\phantom{0000lll0000} & 0.1948\%\phantom{0000lll0000} & 0.328\%\phantom{0000lll000} & 0.552\%\phantom{00000l000} \\
5x Solar Model & 0.1563\%\phantom{0000lll0000} & 0.2170\%\phantom{0000lll0000} & 0.363\%\phantom{0000lll000} & 0.596\%\phantom{00000l000} \\
Slow Rot. Model & 0.1326\%\phantom{000lll00000} & 0.1911\%\phantom{000lll00000} & 0.326\%\phantom{0000lll000} & 0.553\%\phantom{00000l000} \\
Fast Rot. Model & 0.1270\%\phantom{0000lll0000} & 0.1806\%\phantom{0000lll0000} & 0.310\%\phantom{0000lll000} & 0.532\%\phantom{0000l0000} \\
&&&&\\
\emph{Dayside Flux Ratio (Phase$=0.5$)\tablenotemark{a}} &&&\\
1x Solar Model & 0.1194\%\phantom{0000lll0000} & 0.1760\%\phantom{0000lll0000} & 0.301\%\phantom{0000lll000} & 0.523\%\phantom{00000l000} \\
5x Solar Model & 0.1423\%\phantom{0000lll0000} & 0.2035\%\phantom{0000lll0000} & 0.347\%\phantom{0000lll000} & 0.581\%\phantom{00000l000} \\
Slow Rot. Model & 0.1305\%\phantom{000lll00000} & 0.1884\%\phantom{000lll00000} & 0.321\%\phantom{0000lll000} & 0.547\%\phantom{00000l000} \\
Fast Rot. Model & 0.1092\%\phantom{0000lll0000} & 0.1656\%\phantom{0000lll0000} & 0.285\%\phantom{0000lll000} & 0.507\%\phantom{0000l0000} \\
&&&&\\
\emph{Minimum Planet-Star Flux Ratio\tablenotemark{a}} &&&\\
Measured\tablenotemark{b} & $0.0224\%\pm0.0073\%$ & $0.0805\%\pm0.0097\%$ & $<0.311\%$\tablenotemark{c}\phantom{00l00000} & $0.443\%\pm0.027\%$ \\ 
1x Solar Model & 0.0384\%\phantom{0000lll0000} & 0.1063\%\phantom{0000lll0000} & 0.158\%\phantom{0000lll000} & 0.375\%\phantom{0000lll000} \\
5x Solar Model & 0.0382\%\phantom{0000lll0000} & 0.0912\%\phantom{0000lll0000} & 0.150\%\phantom{0000lll000} & 0.359\%\phantom{0000lll000} \\
Slow Rot. Model & 0.0413\%\phantom{0000lll0000} & 0.1124\%\phantom{0000lll0000} & 0.173\%\phantom{0000lll000} & 0.399\%\phantom{0000lll000} \\
Fast Rot. Model & 0.0535\%\phantom{0000lll0000} & 0.1248\%\phantom{00000lll000} & 0.189\%\phantom{0000lll000} & 0.414\%\phantom{0000lll000} \\
&&&&\\
\emph{Nightside Flux Ratio (Phase$=0$)\tablenotemark{a}} &&&\\
1x Solar Model & 0.0605\%\phantom{0000lll0000} & 0.1263\%\phantom{0000lll0000} & 0.194\%\phantom{0000lll000} & 0.413\%\phantom{0000lll000} \\
5x Solar Model & 0.0557\%\phantom{0000lll0000} & 0.1073\%\phantom{0000lll0000} & 0.175\%\phantom{0000lll000} & 0.385\%\phantom{0000lll000} \\
Slow Rot. Model & 0.0424\%\phantom{0000lll0000} & 0.1128\%\phantom{0000lll0000} & 0.174\%\phantom{0000lll000} & 0.400\%\phantom{0000lll000} \\
Fast Rot. Model & 0.0800\%\phantom{0000lll0000} & 0.1423\%\phantom{00000lll000} & 0.229\%\phantom{0000lll000} & 0.450\%\phantom{0000lll000} \\
&&&&\\
\emph{Minimum Flux Offset [h] } &&&\\
Measured &  $-6.43\pm0.82$\phantom{0000l000} &  $-1.37\pm1.00$\phantom{0000l000} & $<10$\phantom{0000l00000000} & $<0$\phantom{0000l00000000}\\
1x Solar Model & $-8.4$\phantom{00000000l000000} & $-8.9$\phantom{000000l00000000} & $-7.8$\phantom{000000000000} & $-7.9$\phantom{000000000000} \\
5x Solar Model & $-6.9$\phantom{00000000l000000} & $-6.9$\phantom{00000000l000000} & $-5.9$\phantom{000000000000} & $-5.9$\phantom{000000000000} \\
Slow Rot. Model & $-2.7$\phantom{0000000l0000000} & $-2.3$\phantom{00000000l000000} & $-2.4$\phantom{000000000000} & $-2.3$\phantom{000000000000} \\
Fast Rot. Model & $-10.6$\phantom{0000000l0000000} & $-10.6$\phantom{00000000l000000} & $-10.1$\phantom{000000000000} & $-10.1$\phantom{000000000000} \\
&&&&\\
\emph{Maximum Flux Offset [h] } &&&\\ 
Measured & $-5.29\pm0.59$\phantom{000lll000}  & $-2.98\pm0.82$\phantom{0000l000}  & $-3.5\pm0.4$\phantom{0000000}  & $-5.5\pm1.2$\phantom{0000000}\\ 
1x Solar Model & $-7.8$\phantom{0000000l0000000} & $-7.6$\phantom{0000000l0000000} & $-7.3$\phantom{000000000000} & $-7.2$\phantom{000000000000} \\
5x Solar Model & $-6.0$\phantom{00000000l000000} & $-5.2$\phantom{0000000l0000000} & $-4.4$\phantom{000000000000} & $-4.0$\phantom{000000000000} \\
Slow Rot. Model & $-2.5$\phantom{00000000l000000} & $-2.7$\phantom{00000000l000000} & $-2.8$\phantom{000000000000} & $-3.1$\phantom{000000000000} \\
Fast Rot. Model & $-9.0$\phantom{00000000l000000} & $-8.6$\phantom{00000000l000000} & $-7.9$\phantom{000000000000} & $-7.6$\phantom{000000000000} \\
\enddata
\tablenotetext{a}{Because our maximum fluxes are located close to a phase of 0.5, we approximate the maximum fluxes as the measured secondary eclipse depths.  The minimum flux is then simply the maximum flux minus the measured phase curve amplitude.  Model fluxes are calculated at a single time corresponding to either the average location of the flux maximum/minimum or an orbital phase of either 0 (night side) or 0.5 (day side).}
\tablenotetext{b}{Fluxes are estimated after removing fitted (3.6 and 4.5~\micron) or predicted (8.0 and 24~\micron) stellar flux variations due to spots.}
\tablenotetext{c}{$2\sigma$ upper limit based on the minimum flux estimate of $0.261\%\pm0.025\%$ from \citet{knutson09a}.}
\end{deluxetable*}

In examining the locations of the local maxima and minima, we find that the slow rotator model is the only one that provides an acceptable fit ($<3\sigma$ disagreement) to the measured offsets in the 4.5, 8.0, and 24~\micron~bands.  The 3.6~\micron~band displays a larger offset that is inconsistent with this model at the $5\sigma$~level, and is instead best-matched by the offset in the $5\times$ solar model.  We note that there are a number of model parameters  in addition to rotation rate that could reduce the offsets of the maxima and minima given the complex interplay of radiative, chemical, and advective timescales.  For example, increasing atmospheric drag in the region near the photosphere would decrease the wind speeds and reduce the corresponding offsets \citep[e.g.,][]{rauscher12}.  By controlling the opacities and therefore the photosphere pressures that contribute flux in a given bandpass, atmospheric chemistry also plays a large role in determining the offsets of the observed maxima and minima.  Our models assume equilibrium chemistry, but it is likely that disequilibrium processes are important for HD~189733b \citep{moses11}.  

\begin{figure}
\epsscale{1.2}
\plotone{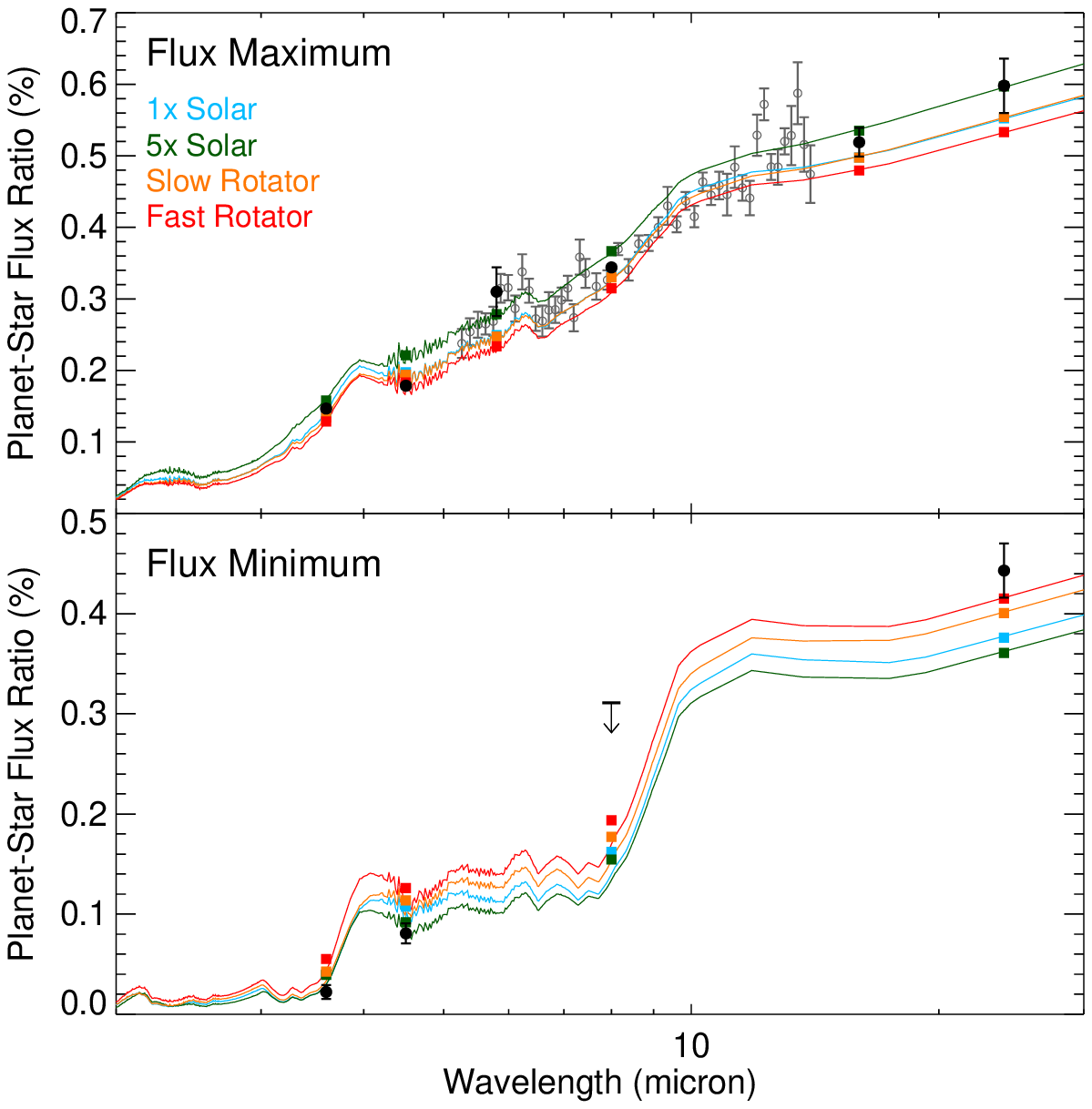}
\caption{Comparison of measured broadband planet-star flux ratios (filled black circles) at 3.6 and 4.5~\micron~(this paper), 5.6, 16, and 24~\micron~\citep{charbonneau08,knutson09a}, and 8~\micron~\citep{knutson09a,agol10}, and an IRS spectrum (open grey circles, Grillmair et al. 2008) vs. 3D model emission spectra at times of minimum and maximum flux, approximately corresponding to the night and day sides of the planet.  Solid lines indicate the model predictions for the $1\times$ solar (blue), $5\times$ solar (green), slow rotator (orange) and fast rotator (red) cases, with the bandpass-integrated fluxes for each model overplotted as filled squares with the same color.}
\label{3D_spectra}
\end{figure}

Although the slow rotator model provides a reasonably good match to the measured locations of the flux maxima and minima, we must next consider whether or not it accurately predicts the planet's emission spectrum at these two points in time.  We find that the our measured maximum fluxes differ by [+11\%, -7\%, +5\%, +8\%] from their predicted values in the [3.6, 4.5, 8.0, 24]~\micron~bands, with a significance of [$3.5\sigma$, $-3.3\sigma$, $4.5\sigma$, $1.2\sigma$], respectively.  On the planet's night side, the minimum fluxes differ by [-46\%, -28\%, +11\%] from their predicted values in the [3.6, 4.5, 24]~\micron~bands, with a corresponding significance of [$-2.7\sigma$, $-3.3\sigma$, $1.6\sigma$], respectively.  The $2\sigma$ upper limit on the 8~\micron~minimum is consistent with all models.  We show the predicted spectra at times of minimum and maximum flux in Fig. \ref{3D_spectra}, and list the corresponding planet-star flux ratios in the \emph{Spitzer} bands in Table \ref{phase_table}.

We next consider whether or not non-equilibrium chemistry might be able to explain the disagreement between the measured and predicted planet-star flux ratios.  The planet's 3.6 and 4.5~\micron~nightside fluxes display the single largest discrepancies, with valuees approximately one third to one half those predicted by our preferred model.  As discussed in \citet{cooper06}, the time scale for converting between CO and methane is long relative to the vertical mixing time scales in these atmospheres, and as a result the relative abundances of these two molecules should reflect their values in the planet's deeper atmosphere ($\sim1-10$~bars) where chemical quenching can occur.  For planets in the temperature regime of HD~189733b, Cooper \& Showman's models suggest that approximately 20\% of the total carbon should exist in the form of methane on both the day and night sides, with the remainder residing primarily in CO.  If we compare this to chemical equilibrium predictions for HD 189733b's night side, we find that this leads to an excess of CO and a deficit of CH$_4$ at the pressures probed by our observations.  Increasing the amount of CO in our nightside model would suppress the 4.5~\micron~flux, bringing it into good agreement with our measured value in this band.  This result is also consistent with our constraints on the planet's transmission spectrum (see \S\ref{depth_discussion}), which suggest the presence of additional CO absorption at the day-night terminator.  Although the 3.6~\micron~nightside flux is also lower than predicted, we note that the slow rotator model provides a poor match to the measured locations of the flux maximum and minimum in this band, and it is therefore not surprising that it would also fail to match the measured flux on the planet's night side.

Although non-equilibrium chemistry offers a possible solution for HD~189733b's night side, the picture on the day side is more mixed.  \citet{moses11} and \citet{visscher11} find that their non-equilibrium models for this planet's day side have more methane and less CO than the equilibrium predictions, as expected in the limit of strong vertical mixing ($K_{zz}=10^{11}$).  However, the net effect of this non-equilibrium chemistry in their models was to decrease the dayside fluxes in the 3.6, 8.0, and 24~\micron~bands, while preserving a nearly identical flux in the 4.5~\micron~band; unfortunately this is the opposite of the change needed in this case.  

\section{Conclusions}

In this paper we present new full-orbit and near-continuous phase curve observations of the hot Jupiter HD~189733b in the 3.6 and 4.5~\micron~\emph{Spitzer} bands, which allow us to characterize the atmospheric circulation patterns and corresponding chemistry of HD~189733b's atmosphere.  These data include one transit and two secondary eclipses in each bandpass, which we use to derive an improved estimate for the planet's orbital ephemeris and wavelength-dependent transmission and emission spectrum in these bands.  We confirm that the planet's 4.5~\micron~transit depth is $3\sigma$ smaller than at 3.6~\micron, consistent with the presence of excess CO at the day-night terminator, although our precision is comparable to that reported by \citet{desert09}.  We obtain improved estimates for HD~189733b's dayside planet-star flux ratio of $0.1466\%\pm0.0040\%$ in the 3.6~\micron~band and $0.1787\%\pm0.0038\%$ in the 4.5~\micron~band, corresponding to brightness temperatures of $1328\pm11$~K and $1192\pm9$~K, respectively; these are the most accurate secondary eclipse depths obtained to date for an extrasolar planet.  Our new 3.6~\micron~secondary eclipse depth is $7.5\sigma$ smaller than the value reported in \citet{charbonneau08}, but we find that the uncertainties in this previous measurement, which assumed Gaussian noise, are likely underestimated.  We conclude that there is no convincing evidence for time variability in the measured secondary eclipse depths or times, consistent with the upper limits derived by \citet{agol10} using a more extensive 8~\micron~data set.    

We combine our new 3.6 and 4.5~\micron~phase curves with previously published observations at 8.0 and 24~\micron~in order to characterize the planet's emission spectrum as a function of orbital phase.  We find that the times of minimum and maximum flux occur several hours earlier than predicted for an atmosphere in radiative equilibrium, consistent with the eastward advection of gas by an equatorial super-rotating jet.  The locations of the flux minima in our new data differ from our previous observations at 8~\micron, and we present new evidence indicating that the flux minimum observed in the 8~\micron~is likely caused by an over-shooting effect in the 8~\micron~array.  We fit the planet's dayside spectrum with a 1D radiative transfer model from \citet{burrows08} where the amount of energy transported to the night side is left as a free parameter and find that the corresponding nightside spectrum is in good agreement with our measured values.  This serves to validate studies \citep[e.g.,][]{cowan11} that seek to constrain circulation patterns on hot Jupiters based on dayside spectra alone.  

We then compare our phase curves to the predictions of 3D general circulation models from \citet{showman09} and find that we require models with either a slower-than-synchronous rotation rate or increased drag at the bottom of the atmosphere in order to match the small measured offsets in the locations of the phase curve maxima and minima at 4.5, 8.0, and 24~\micron.  We also find that HD~189733b's 4.5~\micron~nightside flux is $3.2\sigma$ smaller than predicted by these models, which assume that the chemistry is in local thermal equilibrium.  We conclude that this discrepancy is best-explained by vertical mixing, which should lead to an excess of CO and correspondingly enhanced 4.5~\micron~absorption in this region.  This result is consistent with our constraints on the planet's transmission spectrum, which also suggest excess absorption in the 4.5~\micron~band at the day-night terminator.

Looking ahead, it is clear that the questions regarding atmospheric chemistry and circulation patterns on tidally locked planets will continue to recur as the field shifts towards studies of smaller and more earth-like worlds.  Current studies of the atmospheres of low-mass planets such as GJ~1214b \citep{bean10,bean11,berta11,desert11b} and GJ~436b \citep[e.g.,][]{stevenson10,beaulieu11,knutson11} focus almost exclusively on systems with M star primaries, as the lower stellar effective temperature and smaller stellar radius result in proportionally larger transit and secondary eclipse depths.  For late M stars, the location of the habitable zone is within the region in which we would expect tidal locking to occur \citep{kasting93}; it is therefore likely that the first atmosphere studies of potentially habitable worlds with the \emph{James Webb Space Telescope} will focus on these tidally locked systems.  

\acknowledgments

We would like to thank Josh Carter for his assistance in implementing a wavelet MCMC analysis for these data.  This work is based on observations made with the \emph{Spitzer Space Telescope}, which is operated by the Jet Propulsion Laboratory, California Institute of Technology, under a contract with NASA.  Support for this work was provided by NASA through an award issued by JPL/Caltech.  HAK was supported in part by a fellowship from the Miller Institute for Basic Research in Science.  

\end{document}